\documentclass[12pt,reqno]{amsart}
\usepackage{graphicx,amssymb,amsmath}
\usepackage{amsaddr}
\usepackage[utf8]{inputenc}
\usepackage{epstopdf}
\usepackage{url,hyperref}	
\usepackage{bm}
\usepackage{color}
\usepackage{array}
\usepackage{enumerate}
\usepackage{doi}

\usepackage{dcolumn}
\usepackage{dcolumn}
\usepackage{graphicx}
\usepackage{xcolor}
\usepackage{amsfonts}
\usepackage{amsmath}
\usepackage{amssymb}
\usepackage{amsthm}
\usepackage{bm}
\usepackage{hyperref}
\usepackage{comment}
\usepackage{subfigure}
\usepackage{mathtools}
\hypersetup{colorlinks,
	citecolor=black,
	filecolor=black,
	linkcolor=black,
	urlcolor=black}
\usepackage{epstopdf}
\usepackage{lipsum}
\usepackage{amsopn}

\usepackage{natbib}
\setcitestyle{authoryear,open={(},close={)}}

\newcommand{\ha}{\frac{1}{2}}
\newcommand{\dt}{\,\mathrm{d}t}
\newcommand{\sech}{\operatorname{sech}}
\newcommand{\ud}{\,\mathrm{d}}
\newcommand{\dW}{\ud W}
\newcommand{\dx}{\ud x}
\newcommand{\ds}{\ud s}
\newcommand{\sechsq}[1]{\operatorname{sech}^2\left( #1\right)}
\newcommand{\pdif}[2]{\frac{\partial #1}{\partial #2}}
\newcommand{\inprod}[1]{\left\langle #1 \right\rangle}

\newcommand{\R}{\mathbb{R}}

\newcommand{\uh}{\hat u}

\usepackage[colorinlistoftodos]{todonotes}

\newcounter{mccomment}

\newcounter{gagcomment}

\date{\today}

\title{A collective coordinate framework to study solitary waves in stochastically perturbed Korteweg-de Vries equations}

\author{Madeleine Cartwright}
\email{mcar4493@uni.sydney.edu.au}
\address{School of Mathematics and Statistics, The University of Sydney, Sydney, New South Wales, 2006 Australia}
\author{Georg~A. Gottwald}
\email{georg.gottwald@sydney.edu.au}
 \address{ School of Mathematics and Statistics, The University of Sydney, Sydney, New South Wales, 2006 Australia}

\begin{document}

\maketitle
	
%%%%%%%%%%%%%%%%%%%%%%%%%%%%%%%%%%%%%%%%%%%%%%%%%%%%%%%%%%%%%%%%%
	
\begin{abstract}
Stochastically perturbed Korteweg-de Vries (KdV) equations are widely used to describe the effect of random perturbations on coherent solitary waves. We present a collective coordinate approach to describe the effect on coherent solitary waves in stochastically perturbed KdV equations. The collective coordinate approach allows one to reduce the infinite-dimensional stochastic partial differential equation (SPDE) to a finite-dimensional stochastic differential equation for the amplitude, width and location of the solitary wave. The reduction provides a remarkably good quantitative description of the shape of the solitary waves and its location. Moreover, the collective coordinate framework can be used to estimate the time-scale of validity of stochastically perturbed KdV equations for which they can be used to describe coherent solitary waves. We describe loss of coherence by blow-up as well as by radiation into linear waves. We corroborate our analytical results with numerical simulations of the full SPDE.
\end{abstract}
	
%%%%%%%%%%%%%%%%%%%%%%%%%%%%%%%%%%%%%%%%%%%

\maketitle

%%%%%%%%%%%%%%%%%%%%%%%%%%%%%%%%%%%%%%%%%%%

% REQUIRED
%\begin{keywords}
%	stochastic partial differential equations; stochastic Korteweg-de Vries equation; solitary waves
%\end{keywords}
	
% REQUIRED
%\begin{AMS}
%	60H15, 35Q53, 35A18, 35-XX
%\end{AMS}
	
%%%%%%%%%%%%%%%%%%%%%%%%%%%%%%%%%%%%%%%%%%%%%%%%%%%%%%%%%%%%%%%%%
	
\section{Introduction}
The Korteweg-de Vries equation has been a cornerstone for the description of coherent waves. Originally derived to describe shallow water waves of long wavelength and small amplitude \citep{KdV1895}, it is now used to describe, amongst others waves in plasmas as well as propagation of waves in electrical transmission lines \citep{Crighton1995}. The KdV equation is an integrable equation which supports coherent solitons with particle like behaviour \citep{ZabuskyKruskal65,GardenerGreeneKruskalMiura67}. A natural question to ask is how is this remarkable property of coherence affected by random perturbations? Several stochastically perturbed KdV equations were proposed to model various effects. We consider here stochastically perturbed KdV equations of the form
\begin{align}
\ud u=(6uu_x-u_{xxx})\dt+\sigma R(u)\dW,
\label{e.skdv}
\end{align}
where $W$ denotes Brownian motion. In particular we consider here spatially homogeneous additive perturbations $R(u)=1$ \citep{Wadati83}, fluctuating dissipation $R(u)=u$ leading to space-dependent multiplicative noise \citep{Herman90} and stochastic velocity fluctuations $R(u)=u_x$ in weakly-dispersive environments leading to space-dependent multiplicative noise \citep{Herman90}. In the deterministic case $R(u)=0$ a solution to the KdV equation is the famous one-parameter soliton solution
\begin{align}
\label{e.soliton}
u(x,t)=-2\kappa \sechsq{w(x+\phi)},
\end{align}
with amplitude $\kappa=w^2$ and location $\phi=4w^2 t$ and $w$ controls the width of the soliton. The effect of random perturbations onto such coherent solitary waves in the KdV equation has been studied by means of the inverse scattering transformation \citep{Karpman79,Garnier01}, adiabatic perturbation theory \citep{Herman90} and by collective variable approximations \citep{ArevaloEtAl03}; see also \citep{AbdullaevBook,BassEtAl88,KivsharMalomed89} for a review. We consider here the framework of the collective variable approximation, where the effect of the perturbation is assumed to render the parameters parametrizing the soliton solution time-dependent \citep{Whitham,McLaughlinScott78,AndersonEtAl88,ScottBook}. This reduces an infinite-dimensional stochastic partial differential equation (SPDE) into a finite-dimensional stochastic differential equation (SDE) for the parameters. The collective coordinate approach typically makes use of the geometric structure of the integrable KdV equation by substituting the ansatz into the Lagrangian of the system and assumes that the perturbations vary slowly compared to typical time and spatial scales of the soliton. Here we instead apply the collective coordinate framework developed in \citet{CartwrightGottwald19} which, instead of working within the Lagrangian, views the restriction of the solution $u(x,t)$ to be of a soliton form with time-dependent parameters as a Galerkin approximation, minimizing the error associated with such an ansatz.  
	
The framework developed in \cite{CartwrightGottwald19} was designed to describe travelling waves in dissipative equivariant SPDEs. It relies on a decomposition of the dynamics into the dynamics along the group and the dynamics orthogonal to it. In \cite{CartwrightGottwald19} it was argued that the noise can freely move along the neutrally stable group orbit, implying a Brownian motion of the front interface, whereas it is controlled in the strongly contracting hyperbolic shape dynamics. This argument has since been made rigorous by \cite{HamsterHupkes2020}. This method has also been successfully applied to describe travelling waves in deterministic dissipative partial differential equations \citep{GottwaldKramer04,MenonGottwald05,MenonGottwald07,MenonGottwald09,CoxGottwald06} and to describe the dynamics of deterministic and stochastic phase oscillators \citep{Gottwald15,Gottwald17,HancockGottwald18,WenqiEtAl20,SmithGottwald19,SmithGottwald20}. It is hence interesting to see if the collective coordinate framework, as formulated in \cite{CartwrightGottwald19}, can be applied to conservative SPDEs which lack any hyperbolicity in the shape dynamics, and hence where the noise directly affects the shape parameters.  
	
Perturbations to the KdV equations typically lead to the radiation of linear waves from the soliton which may non-trivially interact with it. This effect is per construction not captured by collective coordinate approaches. Here we introduce an additional perturbative approach to the collective coordinate approach, motivated by our point of view of performing the collective coordinate reduction within a Galerkin approximation framework. This allows us to determine a coherence time of the solution beyond which the solution ceases to have a well-defined coherent shape.  
	
The paper is organised as follows. In Section~\ref{s.cc} we review the framework of stochastic collective coordinates. The following sections are concerned with the various stochastically perturbed KdV equations. Section~\ref{s.add} considers additive noise $R(u)=1$, and we show that our collective coordinate approach reduces to the analytical solution found in \cite{Wadati83}. Section~\ref{s.u} is concerned with the case of fluctuating dissipation $R(u)=u$ and contains an extension of the collective coordinate framework to incorporate, to first order, the effect of radiation. Section~\ref{s.ux} deals with the case of fluctuating velocities $R(u)=u_x$, where we show that collective coordinates accurately describe the blow-up of this ill-posed SPDE, hence providing a time-scale for which this equation may describe the effect of the perturbation on coherent structures. We present numerical results illustrating the ability of our approach to capture the effect of additive and multiplicative noise on the dynamics of solitary waves. We conclude in Section~\ref{s.discussion} with a discussion and an outlook.
	
%%%%%%%%%%%%%%%%%%%%%%%%%%%%%%%%%%%%%%%%%%%%%%%%%%%%%%%%%%%%%%%%%
	
\section{Method of stochastic collective coordinates} 
\label{s.cc}
	
We briefly review the method of stochastic collective coordinates proposed in \cite{CartwrightGottwald19}. We formulate the method for general SPDEs of the form
\begin{align}
\partial_t u(x,t) = F(u) + \dot\eta(u,x,t),
\label{e.SPDEgen}
\end{align}
with noise $\eta(u,x,t)=\sigma R(u)W_t$ with one-dimensional Brownian motion $W_{t}$ and $x\in \Omega$. For the stochastic KdV equation (\ref{e.skdv}) we have $F(u) = 6uu_x-u_{xxx}$. For multi-dimensional noise the reader is referred to \cite{CartwrightGottwald19}. The underlying assumption of collective coordinates is that the solution can be approximated by some ansatz function $\hat u(x,t; \bf{c})$ for some time-dependent, so called collective coordinates ${\bf{c}}\in\R^n$. For the stochastically perturbed KdV equation (\ref{e.skdv}) a natural choice is
\begin{align}
u(x,t) \approx \hat u(x,t;{\bf{c}})=-2\kappa(t) \sechsq{w(t)(x-\phi(t))}+\beta(t)
\label{e.uhat}
\end{align}
with now time-dependent parameters ${\bf{c}}=\{\kappa,w,\phi,\beta\}$. We allow here for a nonvanishing background $\beta(t)$ which will be used for the additive noise $R(u)=1$ in Section~\ref{s.add}. Note that we allow here for all collective coordinates to evolve independently and do not impose any algebraic relationships between them. For general SPDEs, the ansatz function $\hat u(x,t)$ would need to be judiciously chosen to capture the character of the solution of the SPDEs, for example through matching numerical simulations. 
	
The dynamics of the infinite-dimensional SPDE is encoded in the temporal evolution of the finite-dimensional collective coordinates ${\bf{c}}(t)$. We present in this section the derivation in the general form. In the subsequent sections  we then evaluate the resulting evolution equations for our special case of the stochastic KdV equation (\ref{e.skdv}) with $F(u) = 6uu_x-u_{xxx}$ with ansatz solution (\ref{e.uhat}) and ${\bf{c}}=\{\kappa,w,\phi,\beta\}$, and consider several perturbations $R(u)$. We assume that the collective coordinates evolve according to SDEs which we write as
\begin{align}
d{\bf{c}} = {\bf{a}}_{\bf{c}}({\bf{c}}) \, dt + \sigma_{\bf{c}}({\bf{c}}) \, d B(t),
\label{e.cc_ansatz_gen}
\end{align}
where $dB_t$ is one-dimensional Brownian motion. Here the subscripts in the drift terms ${\bf{a}}_{\bf{c}}$ and in the diffusion terms $\sigma_{\bf{c}}({\bf{c}})$ refer to the collective coordinates; i.e., ${\bf{a}}_{c_{j}}$ and $\sigma_{c_{j}}$ denote the drift and diffusion term, respectively, for the collective coordinate $c_j$. Inserting the ansatz function $\hat u(x,t; \bf{c})$ into the SPDE we obtain, upon employing It\^o's formula, the error 
\begin{align*}
d\mathcal{E}(x,t;{\bf{c}})
= \pdif{\uh}{c_j} d c_j + \frac{1}{2} dc_l \pdif{^2\uh}{{c_l} {\partial{c_j}}} dc_j 
- F(\uh) \, dt -\sigma R(\uh) \,  dW(t) ,
\end{align*}
associated with restricting the solution space to the ansatz function (\ref{e.cc_ansatz_gen}) spanned by the collective coordinates $\mathbf{c}$, 
where we used Einstein's summation convention to simplify notation. In the language of Galerkin approximations the error $d\mathcal{E}$ is referred to as residual. Substituting (\ref{e.cc_ansatz_gen}) and collecting only terms up to order $dt$ we obtain, using the independence of the Brownian motion, 
\begin{align*}
d\mathcal{E}(x,t;{\bf{c}}) 
= \left[
\pdif{\uh}{c_j} a_{c_j} + \frac{1}{2} \sigma_{c_l}\pdif{^2\uh}{{c_l} {\partial{c_j}}} \sigma_{c_j} -F(\uh)
\right]\, dt 
+ 
\left[
\pdif{\uh}{c_j} \sigma_{\rm{c_{j}}}\,dB_{t}-\sigma R(\uh) \,  dW(t)
\right] .
\end{align*}
To maximize the degree to which the collective coordinates approximate solutions of the SPDE, we require that the residual $d\mathcal{E}$ does not project onto the subspace spanned by the collective coordinates. Hence we require that the residual $d\mathcal{E}$ lies in the orthogonal complement to the tangent space of the solution manifold spanned by $\pdif{u}{{c_i}}$ , $i=1,\cdots,n$. Projecting the residual eliminates the spatial dependency and we obtain a system of $n$ algebraic equations for the drift and diffusion coefficients, determining the temporal evolution of the collective coordinates. These orthogonality conditions can be separated into terms corresponding to drift and to diffusion, i.e. terms which are multiplied by $dt$ or by $\sqrt{dt}$, respectively. The $n$ drift contributions are given by 
\begin{align}
\langle \frac{\partial \uh}{\partial c_i} \frac{\partial \uh}{\partial c_j}\rangle  a_{c_j} 
+ \frac{1}{2} \sigma_{c_l}\sigma_{c_j}  \langle \frac{\partial \uh}{\partial c_i} \pdif{^2\uh}{{c_l} {\partial{c_j}}}\rangle  
-\langle  \frac{\partial u}{\partial c_i} F(\hat u) \rangle  
&= 0
\label{e.cc_drift_gen}
\end{align}
for $i=1,\cdots,n$ and the $n$ diffusion contributions, which balance the Brownian motion of the SPDE with the Brownian motion of the collective coordinate system, are given by
\begin{align}
\sigma_{c_j}\langle \frac{\partial \uh}{\partial c_i} \pdif{\uh}{c_j} \rangle\,   dB_{t} =  \sigma \langle \frac{\partial \uh}{\partial c_i} R(\hat u) \rangle \,  dW(t)
\label{e.cc_noise_gen}
\end{align}
for $i=1,\cdots,n$. Note that we can (in principle) achieve pathwise approximation of the solutions with $dB_{t} = dW_{t}$. Together with the $n$ equations for the drift coefficients $a_{c_i}$ (\ref{e.cc_drift_gen}) this determines the drift and diffusion coefficients in the evolution equation for the collective coordinates (\ref{e.cc_ansatz_gen}). We evaluate all relevant inner products for the particular case of the ansatz function (\ref{e.uhat}) for the stochastic KdV equation (\ref{e.skdv}) in Appendix~\ref{a.formulae}.\\
	
We remark that our collective coordinate approach is different to the variational Lagrangian approach adopted in \cite{Whitham,AndersonEtAl88,BassEtAl88,KivsharMalomed89}. Whereas therein the variational form of the KdV equation is directly exploited we here view the collective coordinate approach as a Galerkin approximation, minimizing the residual. 
In Appendix~\ref{a.lagrange} we show how the two approaches differ, even in the case of deterministic perturbations. In particular, we illustrate that the additional structure provided by the Lagrangian is beneficial when considering small perturbations, however our approach provides a better approximation for larger perturbations. 
%\mcinline{Side note: the lagrangian formulation leads to a restricition in shape, which must be closer to an unperturbed KdV equation, than we do. Is this similar/in some way related to how collective coordinates works well when there is contraction in shape? It doesn't exist in the SKdV equation as is but viewing it as a perturbation of a lagrangian system there is something similar there.} 
Since stochastic noise introduces with nonvanishing probability large perturbations, our approach is preferable for stochastically perturbed variational SPDEs.\\
	
In the following we apply this general framework to the various stochastic KdV equations introduced in the previous section.

%%%%%%%%%%%%%%%%%%%%%%%%%%%%%%%%%%%%%%%%%%%%%%%%%%%%%%%%%%%%%%%%%
	
\section{KdV equation with spatially homogenous additive noise $R(u)=1$}
\label{s.add}
As the simplest stochastic perturbation of the KdV equation (\ref{e.skdv}),  we consider $R(u)=1$ with 
\begin{equation}
\ud u = (6uu_x-u_{xxx})\dt +\sigma\ud B.
\label{e.skdv_class}
\end{equation}
This SPDE with additive spatially homogeneous noise supports an analytical solution \citep{Wadati83}. Performing a Galilean transformation
\begin{align*}
X=x+m(t)
\end{align*}
with
\begin{align*}
m(t)=6\int_0^tB(s)\ud s,
\end{align*}
where $B(t)=\int_0^t\ud B$, we obtain the deterministic KdV equation
\begin{align*}
U_t-6UU_X+U_{XXX}=0,
\label{e.wtransf}
\end{align*}
for 
\begin{align*}
u(x,t)&=U(X,t)+B(t).
\end{align*}
Hence a solution of (\ref{e.skdv_class}) is given by
\begin{equation}
u(x,t)=-2w^2\sechsq{w(x-\phi(t))+6w\int_0^tW(s)\ud s}+W(t),
\label{e.wsol}
\end{equation}
where $\phi(t)=\phi_{\rm{det}}$ with the location of the unpeturbed KdV soliton 
\begin{align}
\phi_{\rm{det}}=x_0+4w^2t
\label{e.phidet}
\end{align}
for some initial position $x_0$ and parameter $w$.\\
	
We now show that our collective coordinate approach recovers the analytical solution (\ref{e.wsol}). We employ the ansatz solution (\ref{e.uhat}) where the evolution equations (\ref{e.cc_ansatz_gen}) for the collective coordinates ${\bf{c}}=\{\kappa,w,\phi,\beta\}$ are written as
\begin{align*}
\ud\kappa&=a_{\kappa}\dt+\sigma_{\kappa}\dW,\\
\ud w&= a_w\dt+\sigma_w\dW,\\
\ud\phi&=a_{\phi}\dt+\sigma_{\phi}\dW,\\
\ud\beta&=a_{\beta}\dt+\sigma_{\beta}\dW.
\end{align*}
Since $\beta\neq0$, $u$ does not vanish at infinity and we perform integrations over a finite interval of length $2L$ to ensure that $\int\pdif{\uh}{\beta}\beta\dx = \int\beta\dx$ is well defined. Let $\langle \ldots\rangle = \int_{-L}^{L}\ldots\dx$. Then, dropping the hats for ease of exposition, the contributions (\ref{e.cc_noise_gen}) from the diffusion terms are evaluated as
\begin{align}
\left(
\sigma_{\kappa} \inprod{\left(\pdif{u}{\kappa}\right)^2}
+\sigma_w \inprod{\pdif{u}{w}\pdif{u}{\kappa}}
+\sigma_{\beta}\ \inprod{\pdif{u}{\kappa}} 
\right)\, \dW&=
\sigma \inprod{\pdif{u}{\kappa}}\ud B,
\label{e.kstochcont}
\\
\left(
\sigma_{\kappa} \inprod{\pdif{u}{\kappa}\pdif{u}{w}}
+\sigma_{w} \inprod{\left(\pdif{u}{w}\right)^2}
+\sigma_{\beta} \inprod{\pdif{u}{w}}
\right)\, \dW
&=\sigma \inprod{\pdif{u}{w}}\ud B,
\label{e.wstochcont}
\\
\sigma_{\phi} \inprod{\left(\pdif{u}{\phi}\right)^2}\dW
&=0, 
\label{e.phistochcont}
\\
\left(
\sigma_{\kappa} \inprod{\pdif{u}{\kappa}}
+\sigma_w \inprod{\pdif{u}{w}}
+2L\sigma_{\beta}
\right)\, \dW
&=2L\sigma\ud B.
\label{e.betastochcont}
\end{align}
From \eqref{e.phistochcont} we conclude $\sigma_{\phi}=0$. The remaining three equations form a linear system for the remaining three diffusion terms $\sigma_{\kappa,w,\beta}$ which is solved by $\sigma_{\kappa}=\sigma_w=0$ and $\sigma_{\beta} = \sigma$ under the condition of equal noise $\dW=\ud B$.
	
Similarly, upon dropping the hats, we evaluate the contributions (\ref{e.cc_drift_gen}) from the drift terms as
\begin{align*}
\inprod{\left(\pdif{u}{\kappa}\right)^2}a_{\kappa} +\inprod{\pdif{u}{w}\pdif{u}{\kappa}}a_w+\inprod{\pdif{u}{\kappa}}a_{\beta}&=0,
%\label{e.kdetcont}
\\
\inprod{\pdif{u}{\kappa}\pdif{u}{w}}a_{\kappa}+\inprod{\left(\pdif{u}{w}\right)^2}a_w+\inprod{\pdif{u}{w}}a_{\beta}&=0,
%\label{e.wdetcont}
\\
\inprod{\left(\pdif{u}{\phi}\right)^2}a_{\phi}&=	\inprod{\left(6uu_x-u_{xxx}\right)\pdif{u}{\phi}},
%\label{e.phidetcont}
\\
\inprod{\pdif{u}{\kappa}}a_{\kappa}+\inprod{\pdif{u}{w}}a_w+2La_{\beta}&=0.
%\label{e.betadetcont}
\end{align*}
which is solved by $a_{\kappa}=a_w=a_{\beta}=0$ and $a_{\phi}=\inprod{\left(6uu_x-u_{xxx}\right)\pdif{u}{\phi}}/\inprod{\left(\pdif{u}{\phi}\right)^2} = 4w^2$.
	
Performing the limit $L\to\infty$, we recover the exact solution (\ref{e.wsol}) with
\begin{align*}
\phi(t)&=4w_0t-6\int_0^t\beta(s)\ds,\\
\beta(t)&=\sigma W(t),
\end{align*}
and constant $w(t)=w_0$ and constant $\kappa(t)=w_0^2$ for initial position $x_0$ and parameter $w_0$.
	
%%%%%%%%%%%%%%%%%%%%%%%%%%%%%%%%%%%%%%%%%%%%%%%%%%%%%%%%%%%%%%%%%
	
\section{KdV equation with multiplicative noise $R(u)=u$}
\label{s.u}
We now consider the stochastically perturbed KdV equation with multiplicative noise $R(u)=u$,
\begin{align}
\ud u=(6uu_x-u_{xxx})\dt+\sigma u\ud B,
\label{e.u}
\end{align}
which was introduced to study fluctuating damping \citep{Herman90}. We first show that this equation can in fact be transformed into the KdV equation 
\begin{align}
\ud v=\left(6\mu(t)vv_x-v_{xxx}\right)\dt 
\label{e.v}
\end{align}
with time-dependent random coefficient
\begin{align}
\mu(t) = e^{-\frac{\sigma^2}{2}t+\sigma W}
\label{e.mue}
\end{align}
evolving according to geometric Brownian motion. This is achieved by the transformation $v=\mu^{-1}(t)u$ and subsequent application of It\^o's formula with $\ud v=\mu^{-1}\ud u+u\ud \mu^{-1}-\sigma^2\mu^{-1}u\dt$ and $\ud \mu^{-1}=\sigma^2\mu^{-1}\dt -\sigma \mu^{-1}\dW$. The transformation $u=\mu(t)v$ implies that the solution of (\ref{e.u}) involves an overall time-dependent factor of geometric Brownian motion $\mu(t)$. Note that the transformed equation (\ref{e.v}) conserves energy $E_v=\int v^2 dx$. Deterministic KdV equations with time-dependent coefficients as in (\ref{e.v}) are well studied for slowly varying coefficients \citep{KivsharMalomed89,KoKuehl78}. The random coefficient $\mu(t)$, however, is not slowly varying and hence adiabatic perturbation theory cannot be employed here.\\
%, whereas the energy of the untransformed SPDE (\ref{e.u}) is exponentially increasing  in time, caused by the geometric Brownian motion factor $\mu(t)$. 
%\mcinline{The variance of the energy of untransform SPDE increases exponentially in time. However, we know exactly what the energy of $u$ is: $\int u^2\dx = \int (\mu(t) v)^2\dx = \mu^2(t)E_0$.}

\begin{figure}[htbp]
	\centering
	\includegraphics[width = 0.8\columnwidth]{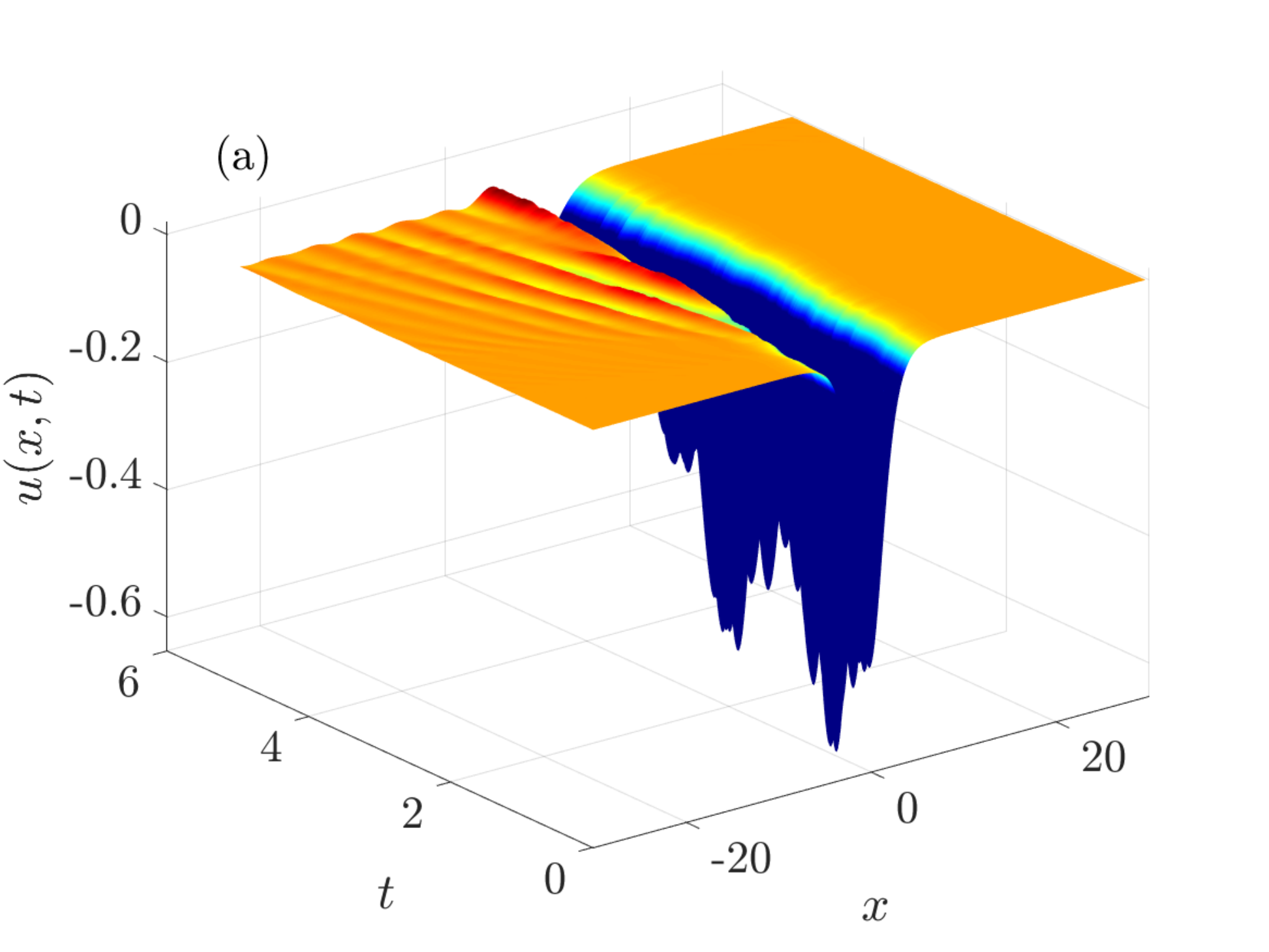}\\
	\includegraphics[width = 0.49\columnwidth]{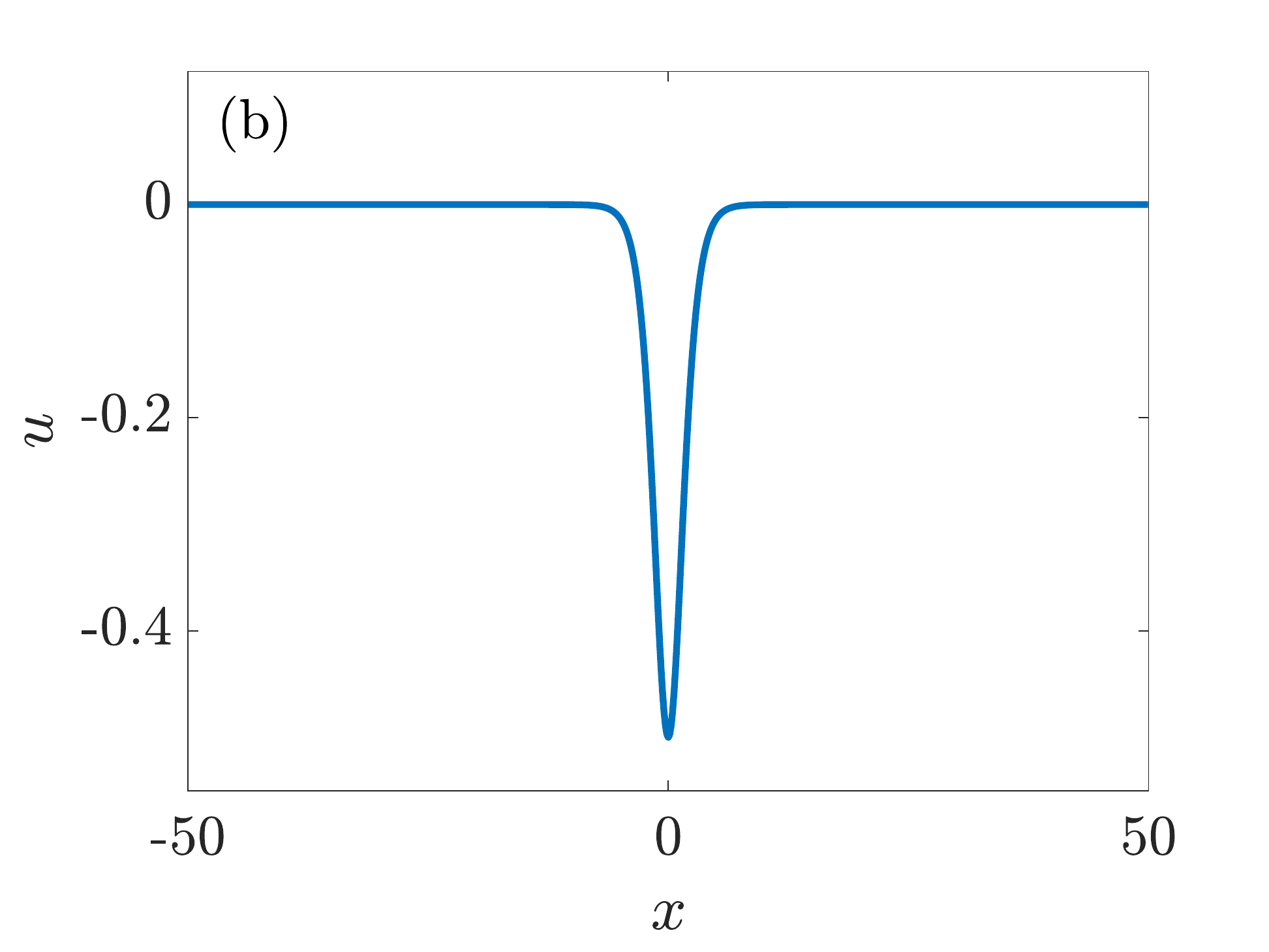}
	\includegraphics[width = 0.49\columnwidth]{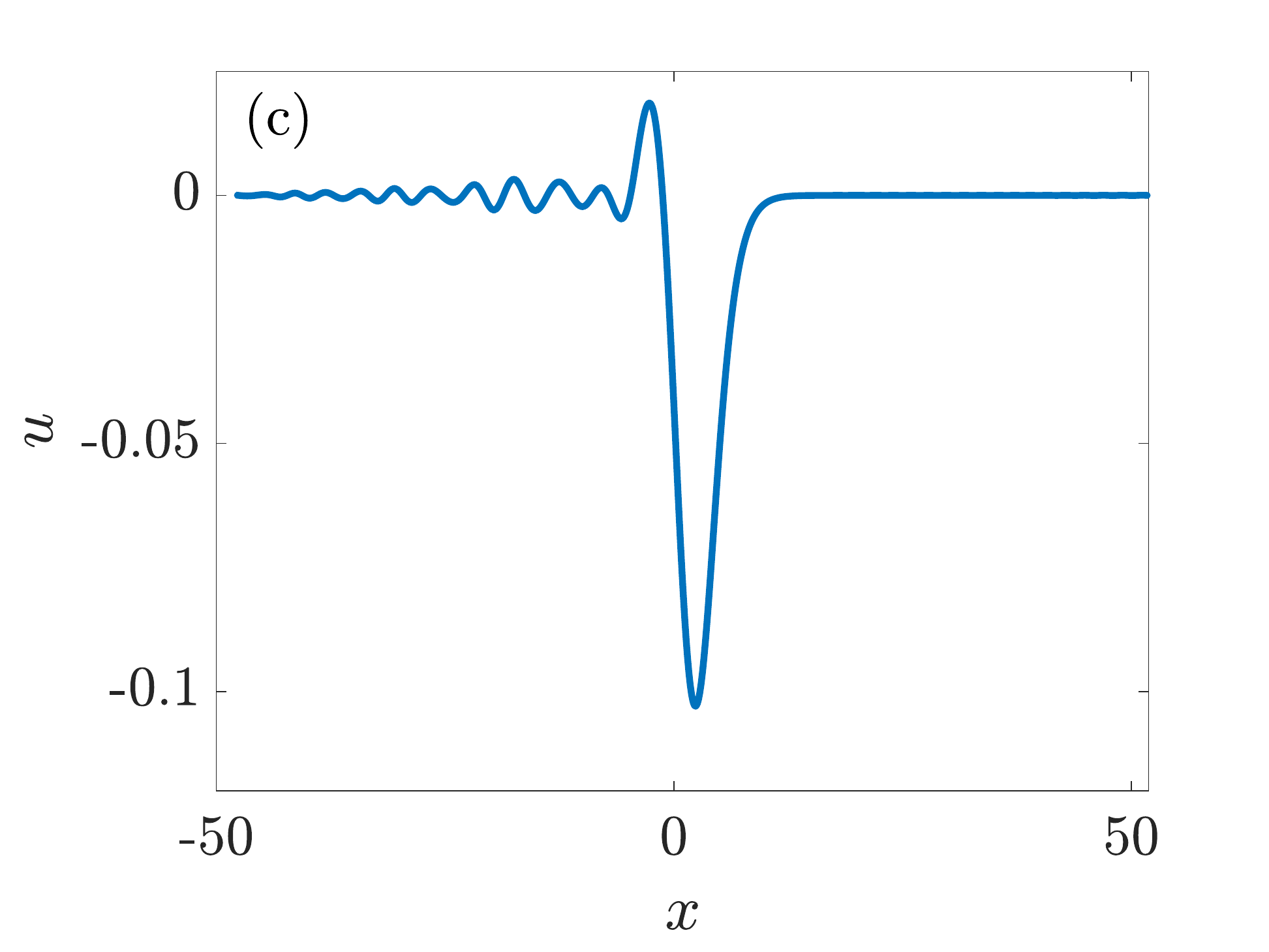}
\caption{(a) Solution of \eqref{e.u} for a fixed realisation of the noise with $\sigma=0.5$. (b) Initial condition $u(x,t=0)$ with $w(0)=0.5$. (c) Snapshot of the solution at $t=5$. Note the different scale.}
\label{f.u_waterfall}
\end{figure}

\begin{figure}[htbp]
	\centering
	\includegraphics[width = 0.8\columnwidth]{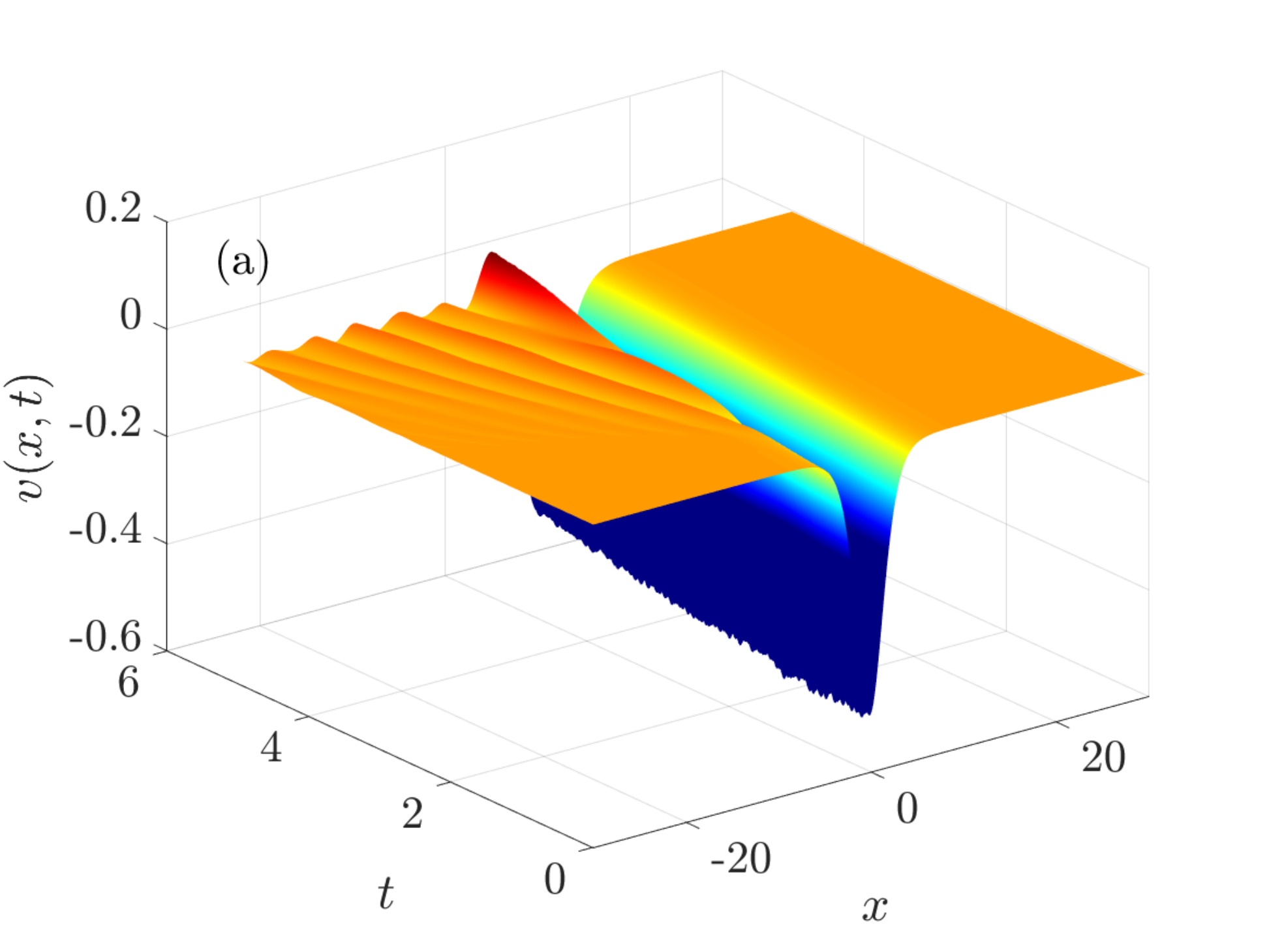}\\
	\includegraphics[width = 0.49\columnwidth]{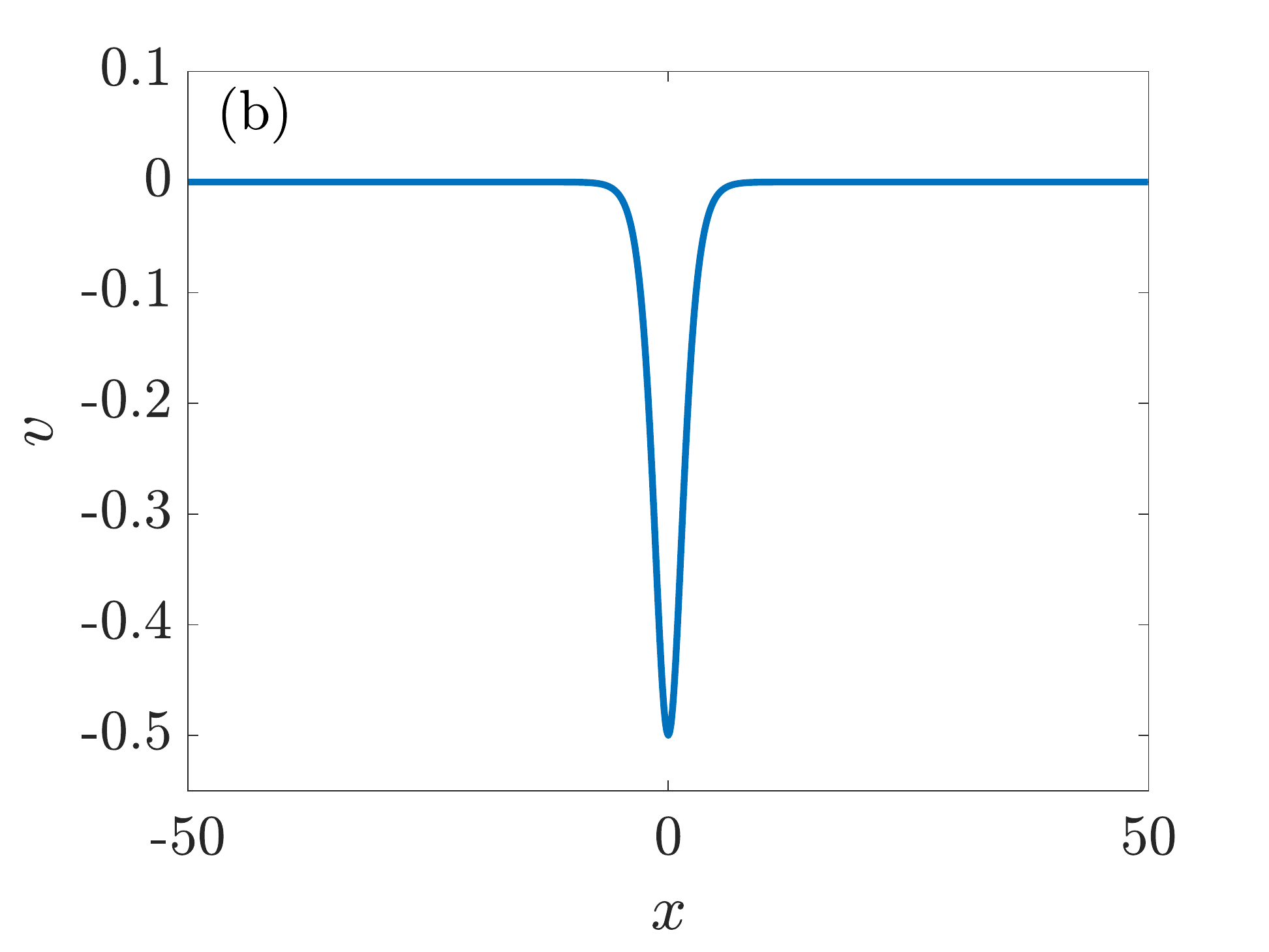}
	\includegraphics[width = 0.49\columnwidth]{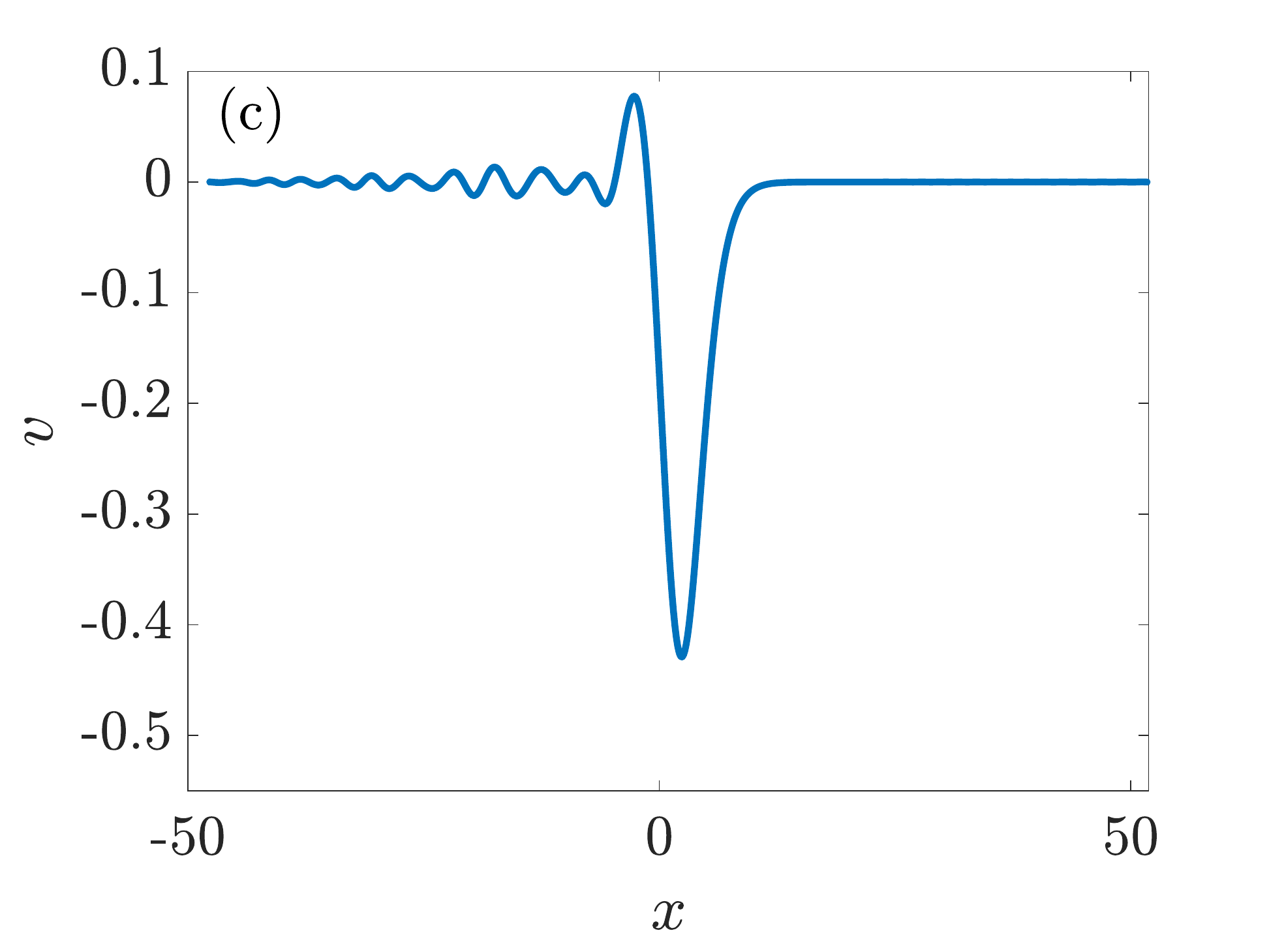}
\caption{(a) Solution of \eqref{e.v} for a fixed realisation of the noise with $\sigma=0.5$. (b) Initial condition $v(x,t=0)$ with $w(0)=0.5$. (c) Snapshot of the solution at $t=5$. }
\label{f.v_waterfall}
\end{figure}

The SPDE (\ref{e.u}) is solved numerically in the spatial domain by finite differencing with periodic boundary conditions, splitting the deterministic part and the stochastic part. The deterministic part is solved by a Crank-Nicolson method for the linear terms and an Adams-Bashforth discretisation for the nonlinearity. The stochastic term is solved by an Euler-Maruyama step \citep{Lord}. For further details of the numerical scheme see Appendix~\ref{a.num}. We choose here a spatial discretisation step of $\Delta x=0.15$ and a temporal integration step $\Delta t = 5\cdot10^{-4}$. 
%We show both the solution $u$ as well as $v=u/\mu(t)$ to eliminate the overall geometric Brownian motion. 
Figure~\ref{f.u_waterfall} shows the time evolution of the solution $u(x,t)$ for the stochastically perturbed KdV equation (\ref{e.u}) with $\sigma=0.5$ for a soliton solution (\ref{e.soliton}) with $w=0.5$ at time $t=0$. Figure~\ref{f.v_waterfall} shows the same for the solution $v(x,t)=u(x,t)/\mu(t)$ of the transformed KdV equation with random coefficient (\ref{e.v}). It is seen that the initial solitary wave disintegrates into radiation and loses coherence. The energy of the solitary waves is pumped into the radiation field. This exchange of energy and the interaction with a radiation field cannot be described by standard collective coordinate approaches which only capture the coherent part. The coherent part, however, as seen in Figure~\ref{f.u_waterfall} becomes less dominant in time. We now present results of our collective coordinate approach and show how we can incorporate the effect of radiation to estimate the time when the solitary wave loses coherence and ceases to be well approximated by the ansatz solution (\ref{e.uhat}) and the collective coordinates ${\bf{c}}=\{\kappa,w,\phi\}$.\\
	
%\begin{figure}[htbp]
%	\centering
%	\includegraphics[width = 0.7\columnwidth]{u_energy_fit.pdf}
%	\caption{Potentially: Energy, normalised by $\kappa_0^2/\kappa^2$. Blue: energy of full solution. Red: energy of fitted soliton}
%	\label{f.u_energy}
%\end{figure}
	
We again seek the temporal evolution for the collective coordinates ${\bf{c}}=\{\kappa,w,\phi\}$ which we write as
\begin{align*}
\ud \kappa&=a_\kappa\dt + \sigma_\kappa\dW,\\
\ud w&= a_{w}\dt+\sigma_w\dW,\\
\ud \phi&=a_\phi\dt+\sigma_\phi\dW.
\end{align*}
The projection of the residual as described in Section~\ref{s.cc} leads to the drift contributions of $\mathcal{O}(dt)$
\begin{align}
2 a_{\kappa}-\frac{\kappa }{w}a_w&=0,
\label{e.drift_u1}
\\
-\frac{\kappa }{w}a_{\kappa} + 2a_w&=0,
\label{e.drift_u2}
\\
a_{\phi}&=-\frac{4}{7}\left(5 w^2-12 \kappa \right),
\label{e.drift_u3}
\end{align}
and to the diffusion contributions of $\mathcal{O}(\sqrt{dt})$
\begin{align*}
\left(2\sigma_{\kappa} - \frac{\kappa }{w}\sigma_w \right)\dW&=2\kappa \sigma\ud B,\\
\left(-\frac{\kappa}{w}\sigma_{\kappa} + 2 \sigma_w \right) \dW&=-\frac{\kappa ^2}{w}\sigma\ud B,\\
\kappa ^2 w\sigma_{\phi}\dW&=0.
\end{align*}
This can be solved for $\dW=\ud B$ to yield
\begin{align}
\ud\kappa &= \sigma \kappa \dW,\\
\ud w&=0,\\
\ud\phi &= \frac{4}{7}\left(12\kappa-5w^2\right) \dt,
\label{e.u_dphi}
\end{align}
which can be analytically solved with
\begin{align}
\kappa(t) &= \mu(t) w^2,
\label{e.u_cc1}
\\
%\phi(t)& = \phi_{\rm{det}}(t) +\frac{48}{7}w^2\int_0^t\left(e^{-\frac{\sigma^2}{2}s+\sigma W_s}-1\right)\ud s,
\phi(t)& = \phi_{\rm{det}}(t) +\frac{48}{7}w^2\int_0^t\left(\mu(s)-1\right)\ud s,
\label{e.u_cc2}
\\
w(t)&=w_0,
\label{e.u_cc3}
\end{align}
for parameter $w_0$, $\mu(t)$ given by (\ref{e.mue}) and the location of the unperturbed deterministic soliton $\phi_{\rm{det}}(t)=4w^2t$ (see (\ref{e.phidet})). Hence, our collective coordinate approach captures the overall amplitude factor $\mu(t)$ of the geometric Brownian motion. We remark that for this it is necessary to allow for the amplitude $\kappa$ and the inverse width $w$ to evolve independently, rather than by requiring $\kappa=w^2$ as implied by collective coordinate approaches relying on a Lagrangian formulation of the KdV equation (see Appendix~\ref{a.lagrange}). However, the shape is solitonic at all times with $\kappa(t)=\kappa_0=w_0^2$ in expectation (note that $\mathbb{E}\mu(t)=1$). Moreover, the collective coordinate approximation (\ref{e.u_cc1})--(\ref{e.u_cc3}) suggests that the noise affects only the position $\phi(t)$; the width remains constant and the amplitude only contains a scaling of the geometric Brownian motion $\mu(t)$.\\
%Furthermore, considering the collective coordinate approximation $v=u/\mu(t)$, which is scaled by the overall geometric Brownian motion, the solution of \eqref{e.v}, we have that both the width and the amplitude of the soliton remain unchanged. In this case the shape remains that of an unperturbed soliton, with only the position $\phi$ being affected by the noise. \\ 
	
\begin{figure}[htbp]
	\centering
	\includegraphics[width = 0.325\columnwidth]{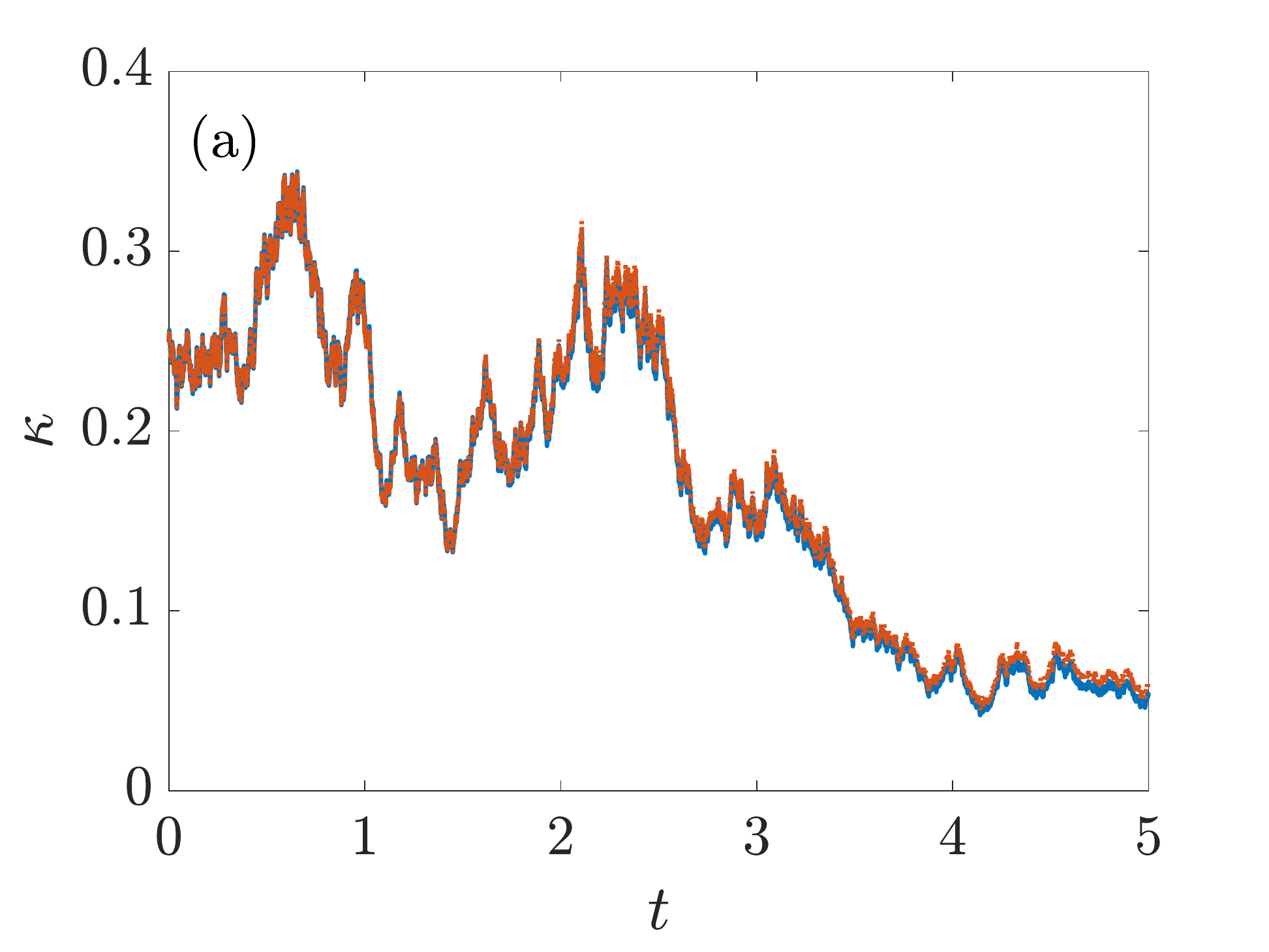}
	\includegraphics[width = 0.325\columnwidth]{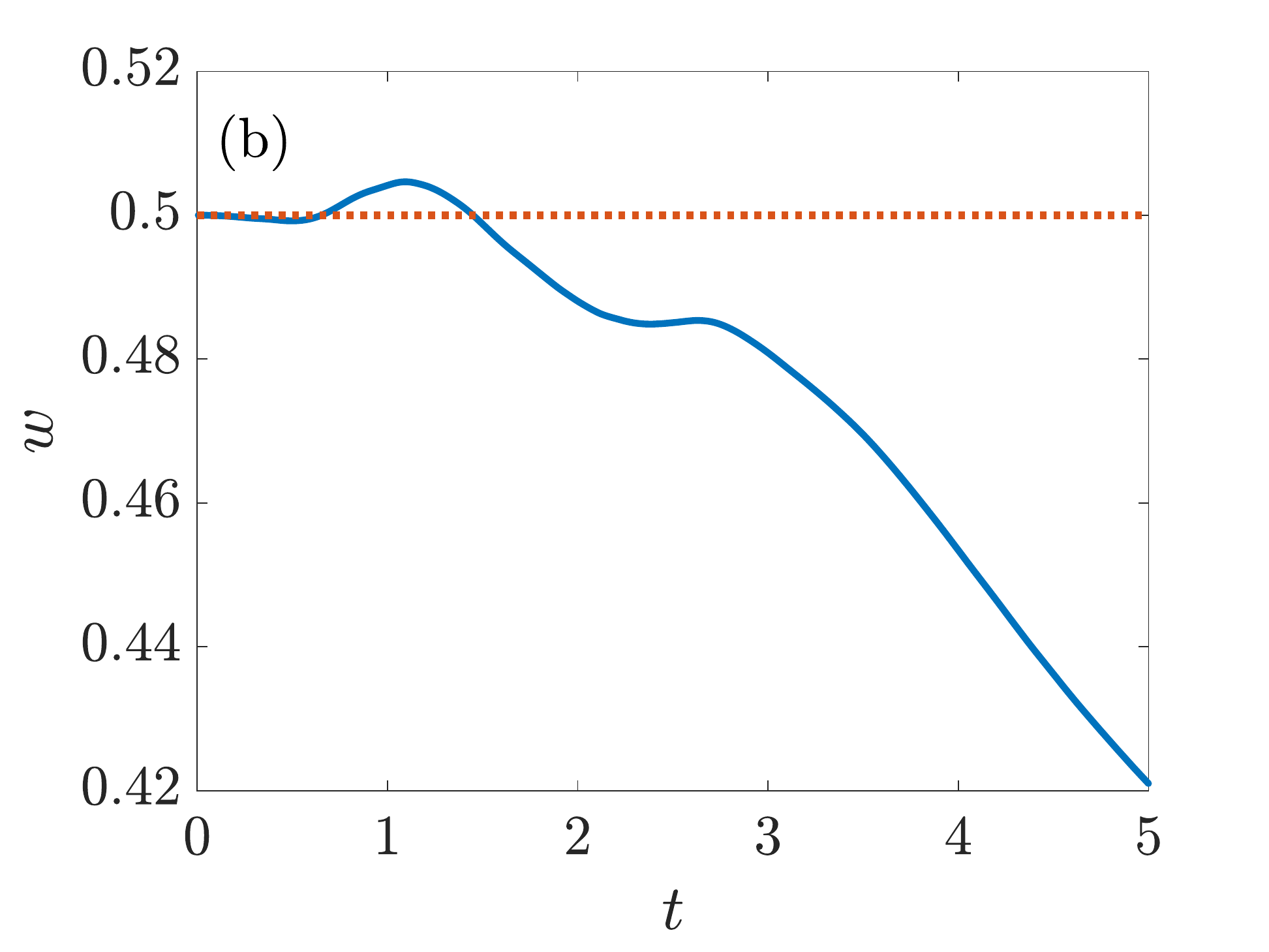}
	\includegraphics[width = 0.325\columnwidth]{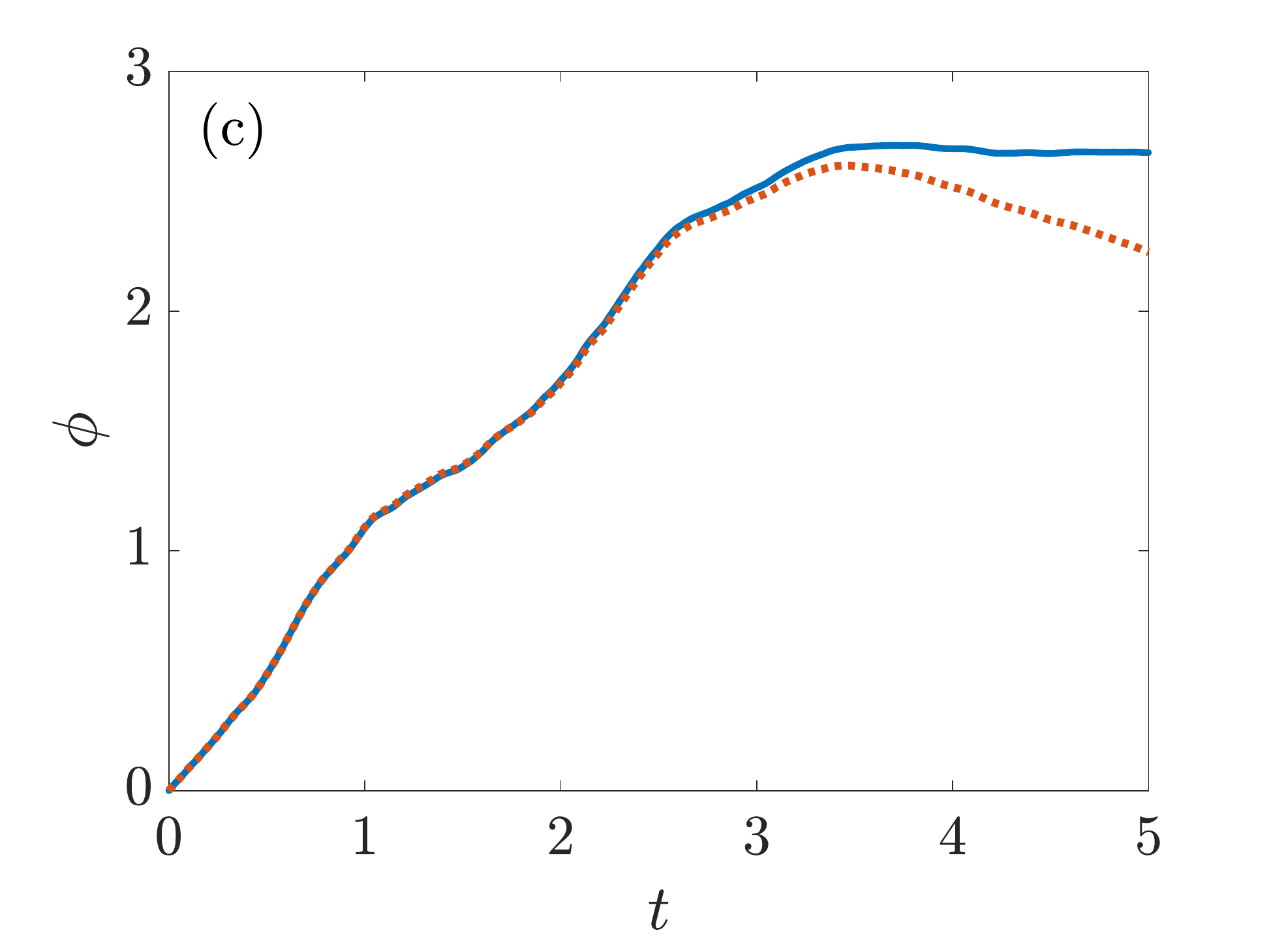}\\
\caption{Amplitude $\kappa$, inverse width $w$ and location $\phi$ of a solitary wave ansatz function (\ref{e.uhat}) for the stochastically perturbed KdV equation (\ref{e.u}) as a function of time. Continuous lines (online blue) are obtained from a direct simulation of (\ref{e.u}). The dotted lines (online red) are the results from the collective coordinate approach (\ref{e.u_cc1})--(\ref{e.u_cc3}). 
%The vertical bars are at $t=???$. 
%for which the absolute error of $w$ is $2\%$. 
Parameters as in Fig~\ref{f.u_waterfall}.}
\label{f.u_cc_t}
\end{figure}

Figure~\ref{f.u_cc_t} shows a comparison of our collective coordinate approach (\ref{e.u_cc1})--(\ref{e.u_cc3}) with a numerical simulation for the stochastically perturbed KdV equation (\ref{e.u}) with $\sigma=0.5$ for one realisation of the noise. To extract the values for the collective coordinates from the direct numerical simulation of the SPDE we perform a nonlinear least square fit to the ansatz solution (\ref{e.uhat}). We remark that the nonlinear least-square fit involves a nonconvex optimisation problem and it is not guaranteed that the fitted collective coordinates correspond to a global minimum. We safeguard against this problem by using the solutions from the previous time-step as initial guess for the optimisation, noting that at $t=0$ we start with a solitary wave and hence the collective coordinates are indeed the global minimum at initial time $t=0$. We see that the collective coordinate approach yields a remarkably good approximation for some time until it deteriorates after $t\approx1.5$. The deterioration is first seen in the inverse width $w$. The amplitude $\kappa$ is dominated by the geometric Brownian motion $\mu(t)$ which forms the overall factor for both the solution of the stochastic KdV equation (\ref{e.u}) and the collective coordinate solution (\ref{e.u_cc1}). Note that at $t=5$ the solitary wave has significantly lost coherence by radiation (cf. Figure~\ref{f.u_waterfall}). As discussed above the loss of coherence is caused by the solitary wave pumping energy into the radiation field and then strongly interacting with it. We now describe how to estimate the coherence time above which the solution ceases to be described by the ansatz function (\ref{e.uhat}), or in other words the time for which an initially coherent solitary wave remains coherent such that it can be captured by our collective coordinate reduction (\ref{e.u_cc1})--(\ref{e.u_cc3}).\\
	
\begin{figure}[htbp]
	\centering
	\includegraphics[width = 0.32\columnwidth]{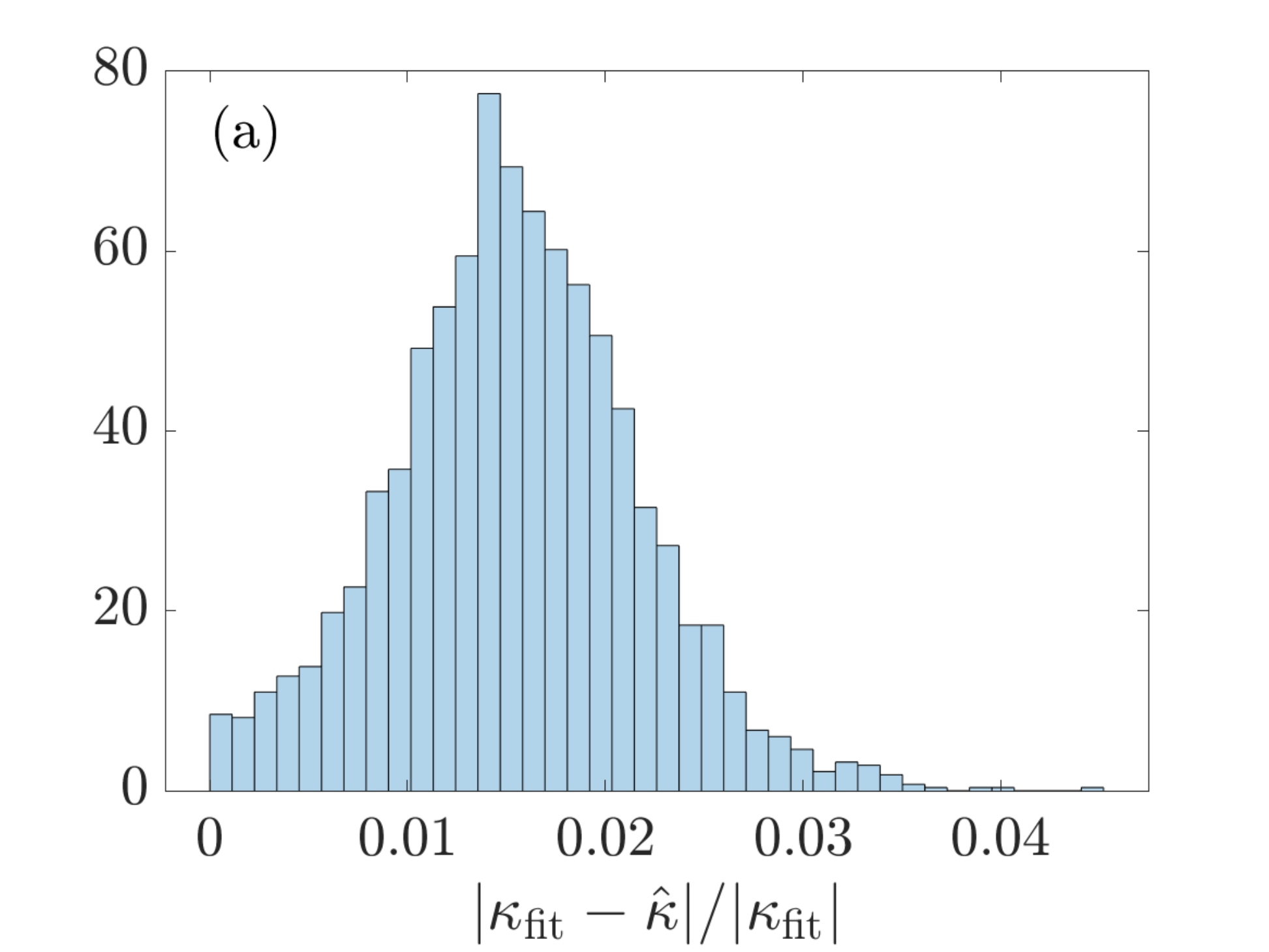}
	\includegraphics[width = 0.32\columnwidth]{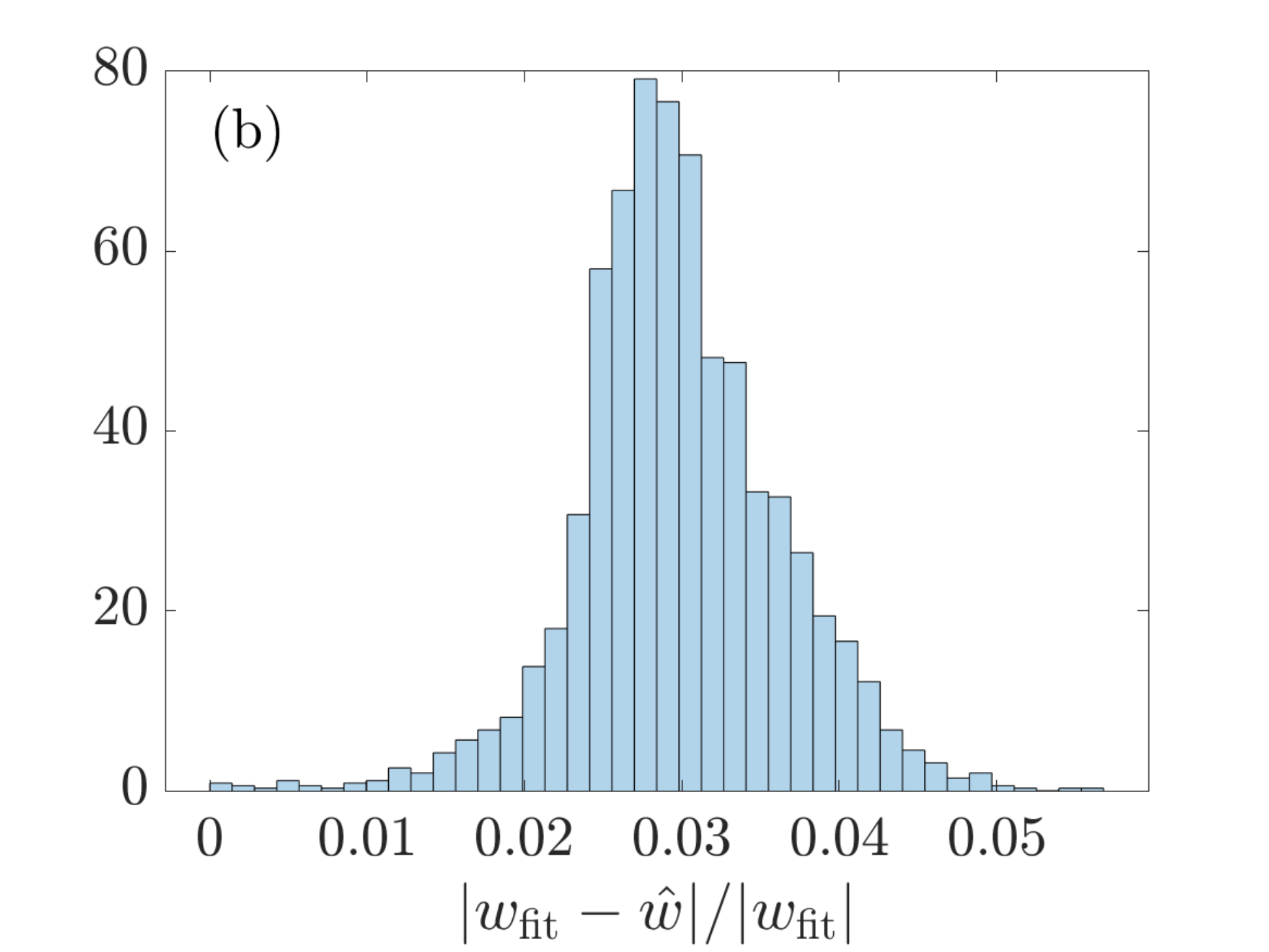}
	\includegraphics[width = 0.32\columnwidth]{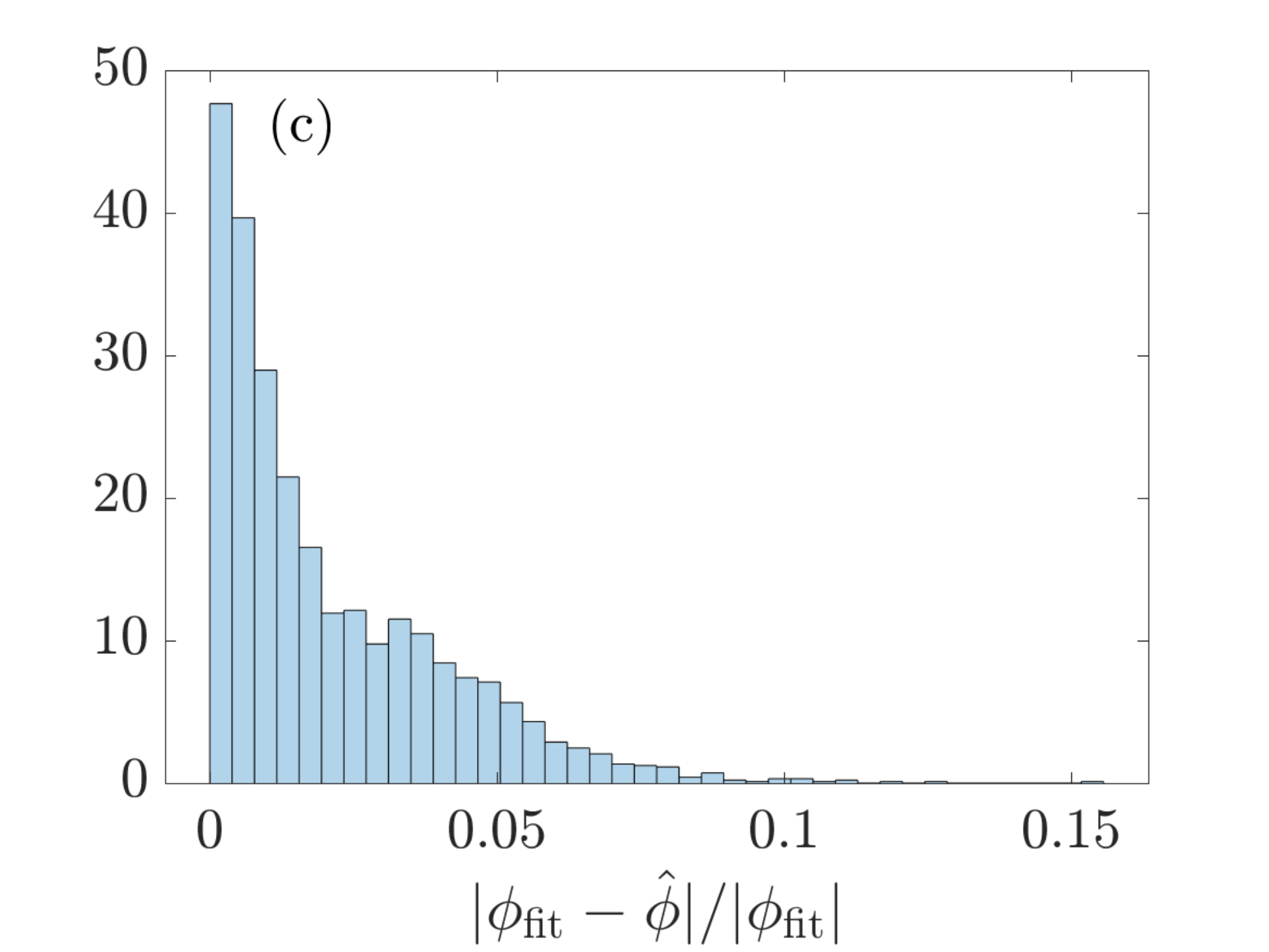}
\caption{Empirical histograms of the errors of the collective coordinates $\kappa$, $w$ and $\phi$ compared to the values obtained by a nonlinear least square fit to (\ref{e.uhat}) for the solution of the full stochastically perturbed KdV equation (\ref{e.u}) at time $\tau_c$ determined by (\ref{e.tauc}) using $\zeta=\tfrac{1}{4}$. Parameters as in Fig~\ref{f.u_waterfall} and the histogram was obtained from $2500$ realisations.}
\label{f.u_delta_cc}
\end{figure}

We first note that the collective coordinate system reproduces the energy $E=\int u^2 dx=\mu^2(t)E_v$ exactly. However, whereas in the full SPDE (\ref{e.u}) this energy is pumped from the solitary wave into the radiation field, this energy is assumed to remain contained within the solitary wave $\hat u(x,t)=\hat{u}(\kappa,w,t)$. We hence need to expand our ansatz function to allow energy to flow outside of the solitary wave. Linearising the unperturbed KdV equation around the solitary wave suggests that we consider as ansatz function
\begin{align}
\tilde u(x,t;{\bf{c}})=\hat u(x,t) + \alpha {\hat u}_x(x,t)
\label{e.uhatx}
\end{align}
with collective coordinates ${\bf{c}}=\{\kappa,w,\phi,\alpha\}$. The correction $\alpha {\hat u}_x$ can be viewed as a term coming from a Taylor expansion around the location $\phi$ of a solitary wave which we write as $\alpha(t) \sim \delta(t)/w$ with
%\gaginline{Is $\alpha \sim \delta$ correct?}
%\mcinline{For small enough $|\alpha|$, yes}
\begin{align}
\delta(t) = w |\phi(t)-\phi_{\rm{det}}(t)|
= \frac{48}{7}w\left|\int_0^t\left(\kappa(t)-w^2\right)\ud s\right|,
\label{e.delta}
\end{align}
where we normalised by the characteristic length scale $w^{-1}$ of the solitary wave. We now define a coherence time $\tau_c$ as the time when the perturbation to the solitary wave starts to become dynamically important, i.e. as the time $\tau_c$ such that $\delta(\tau_c)$ exceeds a threshold $\delta_\theta$, which we formalise as
\begin{align}
\tau_c = \underset{t}{{\rm{arg\, min}}} |\delta(t)-\delta_\theta|.
\label{e.tauc}
\end{align}
To set the threshold $\delta_\theta$ we first define a natural length scale $\lambda$ of the solitary wave. One choice of a natural length scale is the width of the soliton at half-amplitude $\lambda=1.76/w$. We consider several choices of the threshold in terms of the natural length scale with $\delta_\theta=\zeta \lambda$, and show that this correlates well with the observed breakdown of coherence for several values of $\zeta$. Figure~\ref{f.u_delta_cc} illustrates how the definition of the coherence time  (\ref{e.tauc}), which is defined in terms of the collective coordinates and the arbitrary parameter $\zeta=\tfrac{1}{4}$, translates into the ability of our collective coordinate reduction (\ref{e.u_cc1})--(\ref{e.u_cc3}) to capture the true solution of the stochastically perturbed KdV equation (\ref{e.u}). Figure~\ref{f.u_delta_cc} shows the relative error of the collective coordinate predictions for $\kappa$, $w$ and $\phi$ compared to the values obtained by a nonlinear least square fit to (\ref{e.uhat}) for the solution of (\ref{e.u}) at time $\tau_c$ for $\zeta=\tfrac{1}{4}$. It is seen that the relative errors for the amplitude $\kappa$ and the inverse width $w$ are unimodally distributed around mean values of $1.6\%$ and $3.0\%$, respectively. The error in position is decaying approximately monotonically and has a mean relative error of $2.1\%$. This suggests that the introduced coherence time $\tau_c$ correlates well with the relative error made by the collective coordinate approximation, and hence with the loss of coherence of the solitary wave. Choosing different values of $\zeta$ exhibits similar behaviour, albeit with slightly changed relative mean errors, with larger values of $\zeta$ corresponding to larger relative mean errors.\\ 

\begin{figure}[htbp]
	\centering
	\includegraphics[width = 0.49\columnwidth]{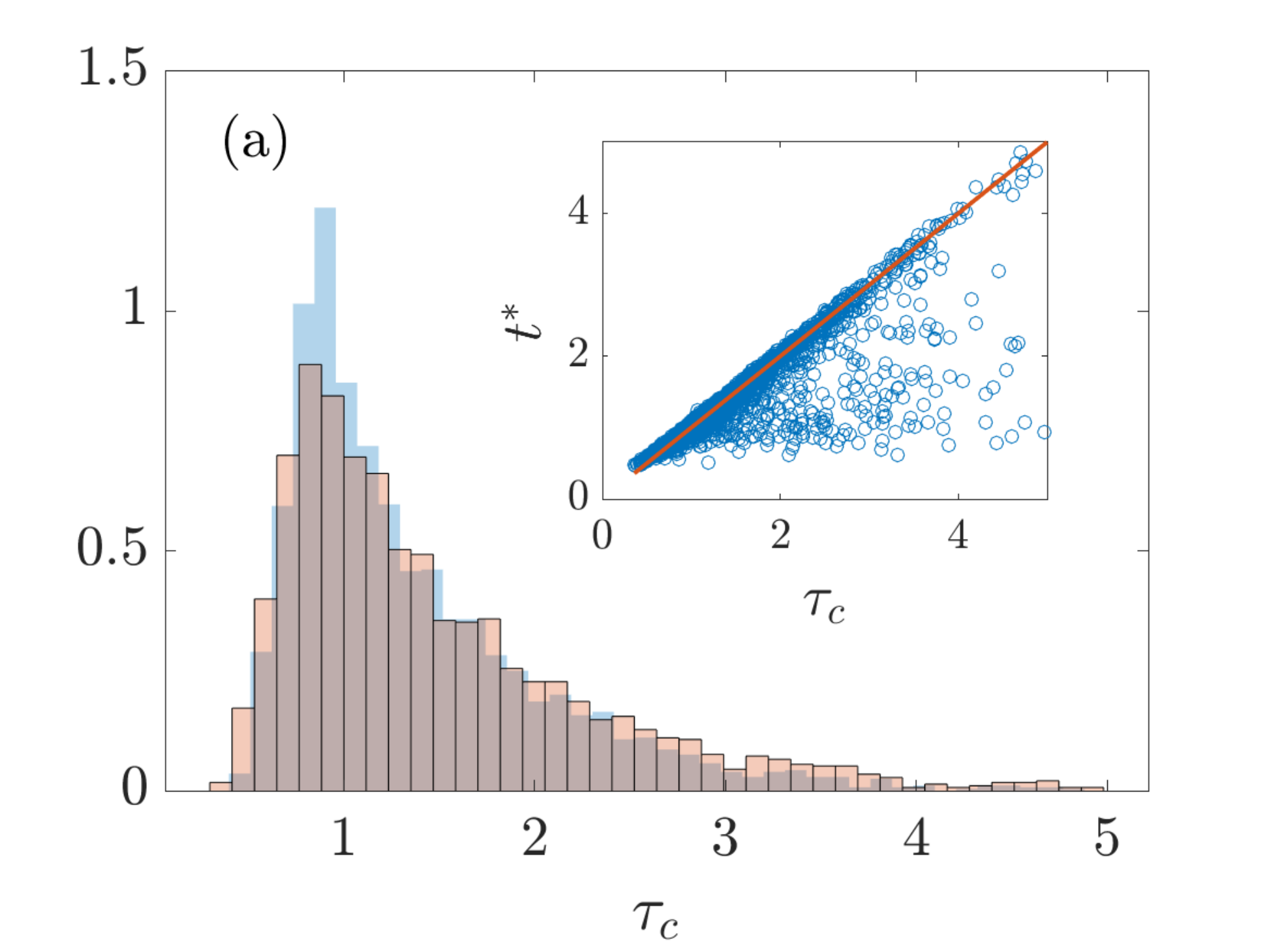}
	\includegraphics[width = 0.49\columnwidth]{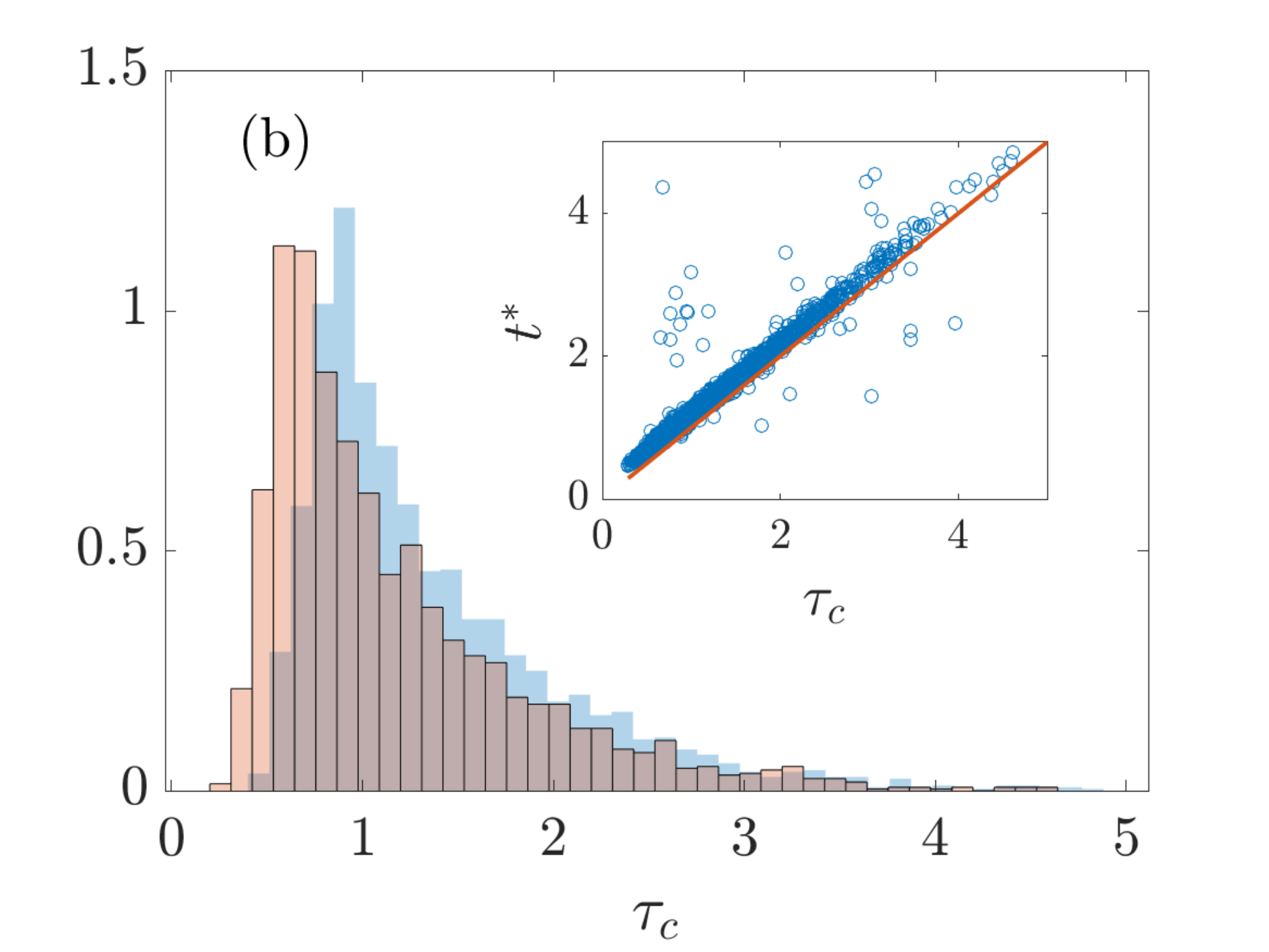}
\caption{Empirical histogram of $\tau_c$ defined in (\ref{e.tauc}) with the empirical histogram of $t^*$ (no outline, online blue) superimposed for comparison for $\zeta=\tfrac{1}{4}$ (left) and $\zeta=\tfrac{1}{6}$ (right). The insets show a scatter plot of $(t^\star, \tau_c)$ together with a reference line $t^*=\tau_c$. Parameters as in Fig~\ref{f.u_waterfall}; the histograms were obtained from $2500$ realisations.} 
%Note that the slight peak at $\tau_c=5$ is due to the simulations being run until $t=5$. }
\label{f.u_tauc}
\end{figure}

To illustrate how the coherence time (\ref{e.tauc}) can be used to estimate the onset of the loss of coherence entirely from information of the collective coordinates, we measure the loss of coherence of the solitary wave solution of the actual SPDE (\ref{e.u}) by recording the loss in shape. As a proxy for the shape we use the inverse width $w$, and define the time $t^\star$ for which the relative error in the inverse width $w$ first exceeds $3.0\%$ using simulations of the actual SPDE (\ref{e.u}). The time $t^\star$ is hence a measure of the coherence time of the actual solitary wave solution. Figure~\ref{f.u_tauc} shows a comparison of the empirical histograms of $\tau_c$ as estimated from collective coordinates using (\ref{e.tauc}) for two values of the free parameter $\zeta$ as well as a histogram of $t^\star$. The two histograms, each obtained from $2,500$ realisations, are remarkably close for both values of $\zeta$. We further show in the insets in Figure~\ref{f.u_tauc} a direct comparison between $t^\star$ and $\tau_c$. We see that most simulations correspond to $t^\star\approx \tau_c$ with a few outliers. This correspondence is only weakly dependent on the choice of the free parameter $\zeta$ (within a reasonable range of  $\zeta\in[1/6,1/3]$). The plots show that collective coordinates are able to provide a reasonable estimate for the loss of coherence of solutions of the SPDE (\ref{e.u}). 

\begin{figure}[htbp]
	\centering
	\includegraphics[width = 0.7\columnwidth]{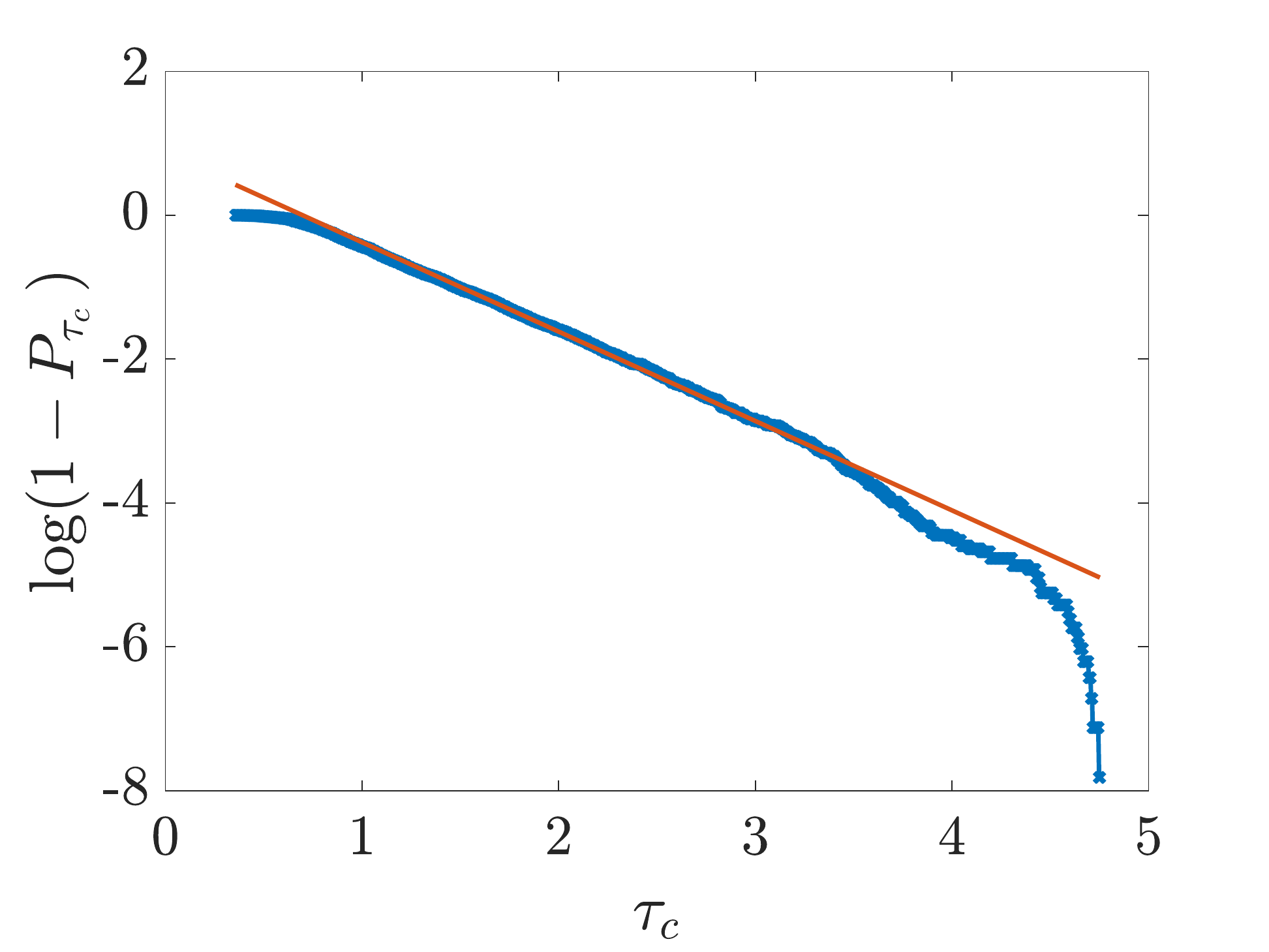}
\caption{Log plot of the empirical normalised cumulative probability density function $p(\tau_c)$ of coherence times calculated from (\ref{e.tauc}). Parameters as in Figure~\ref{f.u_waterfall}. The linear part has a slope of $-1.24$ which is close to the empirical mean of $1.49$.}
\label{f.u_poisson}
\end{figure}

The form of the histogram in Figure~\ref{f.u_tauc} suggests that the break up times are a Poisson process with cumulative probability distribution function
\begin{align*}
P(\tau_c) = 1 - \exp(-\frac{\tau_c}{{\bar{\tau}}_c}),
%\label{e.poisson}
\end{align*}
with mean time of coherence ${\bar \tau}_c$. This is confirmed in Figure~\ref{f.u_poisson}. Linear regression suggests a mean coherence time of $\tau_c=1.30$ which is reasonably close to the empirical mean of the coherence times of $1.49$.\\

We remark that one could perform the collective coordinate approach outlined in Section~\ref{s.cc} for the collective coordinates ${\bf{c}}=\{\kappa,w,\phi,\alpha\}$. One then recovers the expression (\ref{e.delta}) in the limit of small $\alpha$. We present the calculations in Appendix~\ref{a.ccL} together with numerical simulations for completeness.

% up to some constant factors. 
%\gaginline{In what way is (\ref{e.tauc}) recovered?}
%\mcinline{I wouldn't say it's exactly recovered, but we do get $\dot{\alpha}\sim(\kappa-w^2)$. Because of the way the position is split between $\phi$ and $\alpha$ acting as a correction to $\phi$, you can't exactly compare directly between the two.}
	
%\begin{figure}[htbp]
%\centering
%\includegraphics[width = 0.7\columnwidth]{delt_tstar.pdf}
%\caption{Empirical histogram of $\delta(t)$ evaluated at $t=t^\star$ for which the collective coordinate $w$ deviates by $2.5\%$ from the inverse width of the solution of the full stochastically perturbed KdV equation (\ref{e.u}) determined by a nonlinear least square fit to (\ref{e.uhat}). Parameters as in Fig~\ref{f.u_waterfall} and the histogram was obtained from $2500$ realisations.}
%\label{f.u_deltac}
%\end{figure}	
	
%\begin{figure}[htbp]
%	\centering
%	\includegraphics[width = 0.7\columnwidth]{tstar_tauc_hist.pdf}
%\caption{Empirical histogram of $\tau_c$ defined in (\ref{e.tauc}). Parameters as in Fig~\ref{f.u_waterfall} and the histogram was obtained from $2500$ realisations. Note that the slight peak at $\tau_c=5$ is due to the simulations being run until $t=5$. Empirical histogram of $t^*$ (no outline, online blue) is superimposed for comparison.}
%\label{f.u_tauc}
%\end{figure}	
	
%%%%%%%%%%%%%%%%%%%%%%%%%%%%%%%%%%%%%%%%%%%%%%%%%%%%%%%%%%%%%%%%%
	
\section{KdV equation with multiplicative noise $R(u)=u_x$}
\label{s.ux}
We now consider the stochastically perturbed KdV equation with multiplicative noise $R(u)=u_x$,
\begin{align}
\ud u=(6uu_x-u_{xxx})\dt+\sigma u_x\ud B,
\label{e.ux}
\end{align}
which was introduced to study fluctuating velocities \citep{Herman90}. This equation is ill-posed and solutions blow up in time. This blow-up can be readily understood by applying the Galilean transformation $U = u(X,t)$ with $X=x+\sigma B(t)$, which leads to the unstable deterministically perturbed KdV equation with negative diffusion
\begin{align}
U_t=6UU_X-U_{XXX}-\ha\sigma^2U_{XX}.
\label{e.ux2}
\end{align}
The energy 
\begin{align*}
E(t)=\int U^2(X,t)dx 
\end{align*}
grows according to
\begin{align*}
\frac{d}{dt}E = \sigma^2 \int u_x^2\dx.
\end{align*}

Despite this blow up, the stochastically perturbed KdV equation (\ref{e.ux}) has been used to study solitary waves in random environments with fluctuation dissipation \citep{Herman90,BassEtAl88}, often in situations where the noise is spatially confined with $\sigma=\sigma(x)$ \citep{LinEtAl06}. If the spatial extent of the region in which $\sigma(x)\neq 0$ is sufficiently small such that the time of travel of a coherent solitary wave through the fluctuating environment is smaller than the time to develop the instability, equation (\ref{e.ux}) may still be used to model the effect of the random fluctuations on the coherent wave, despite being ill-posed. We shall use collective coordinates to provide an estimate for the time we expect the solitary wave to remain coherent and not blow up. This may serve as a rough guide to modellers to determine the range of validity of their unstable model.\\ 

We remark that by adding diffusion to (\ref{e.ux}) as in
\begin{align*}
u_t=(6uu_x-u_{xxx}+\sigma u_x+\gamma u_{xx})\dt+\sigma u_x\ud B
\end{align*}
one may obtain for $\gamma=\sigma^2/2$ the integrable deterministic KdV equation after applying the Galilean transformation. For  $\gamma=\sigma^2/2$ solutions then inherit the constant shape of the deterministic soliton but experience Brownian motion in their position. For $\gamma>\sigma^2/2$ solutions will experience decay in energy.\\
	
\begin{figure}[htbp]
	\centering
	\includegraphics[width = 0.8\columnwidth]{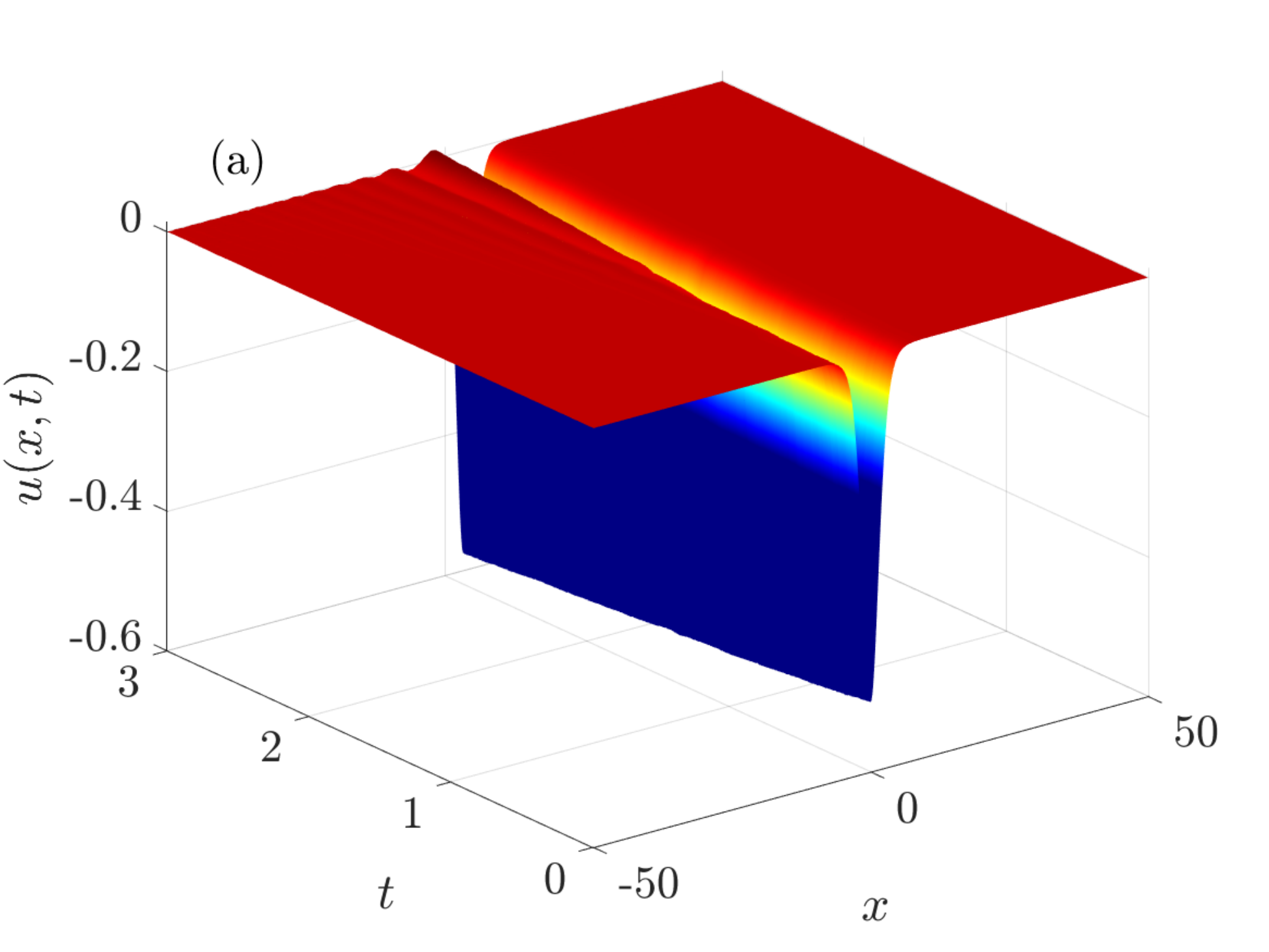}\\
	\includegraphics[width = 0.45\columnwidth]{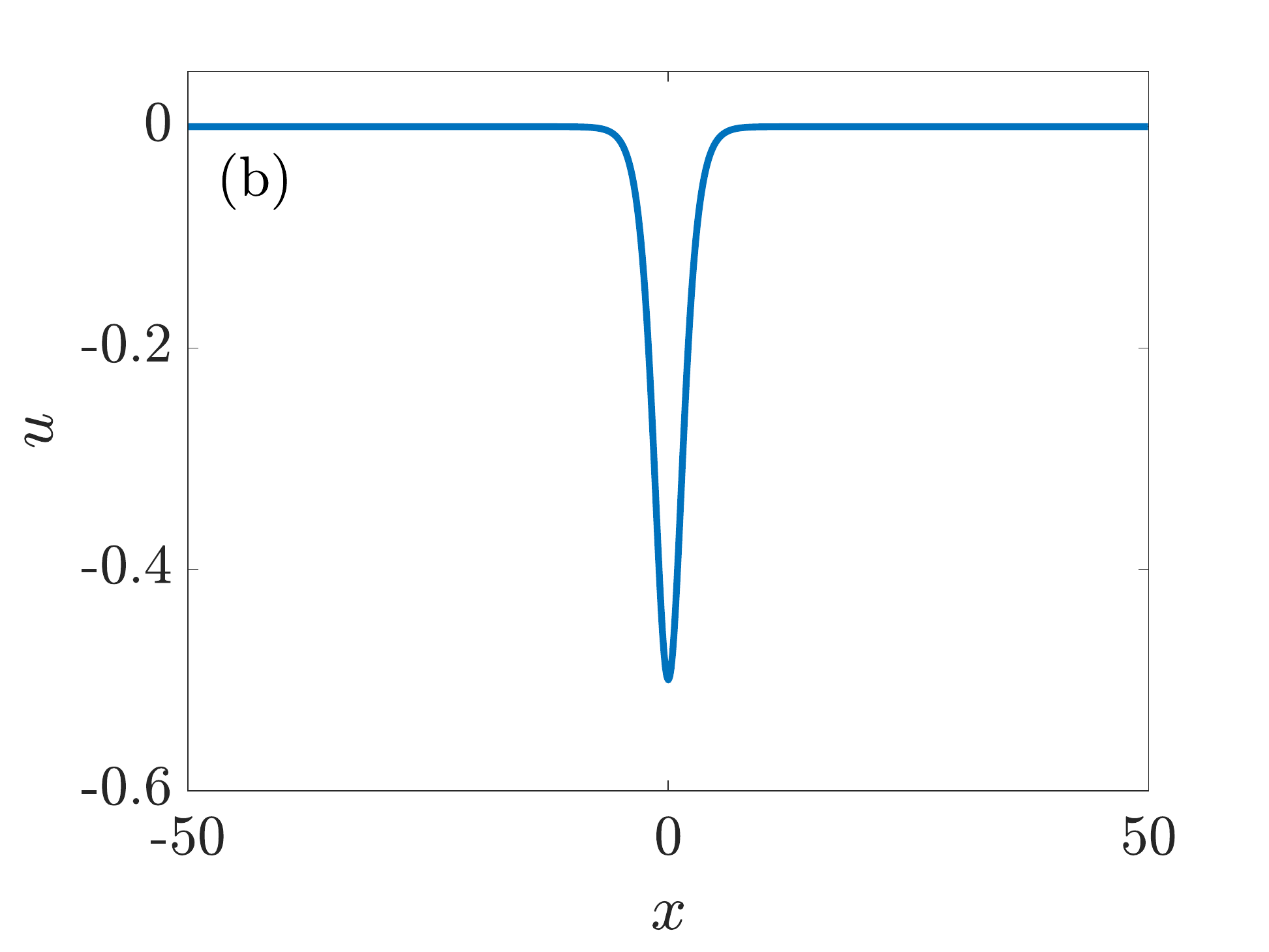}
	\includegraphics[width = 0.45\columnwidth]{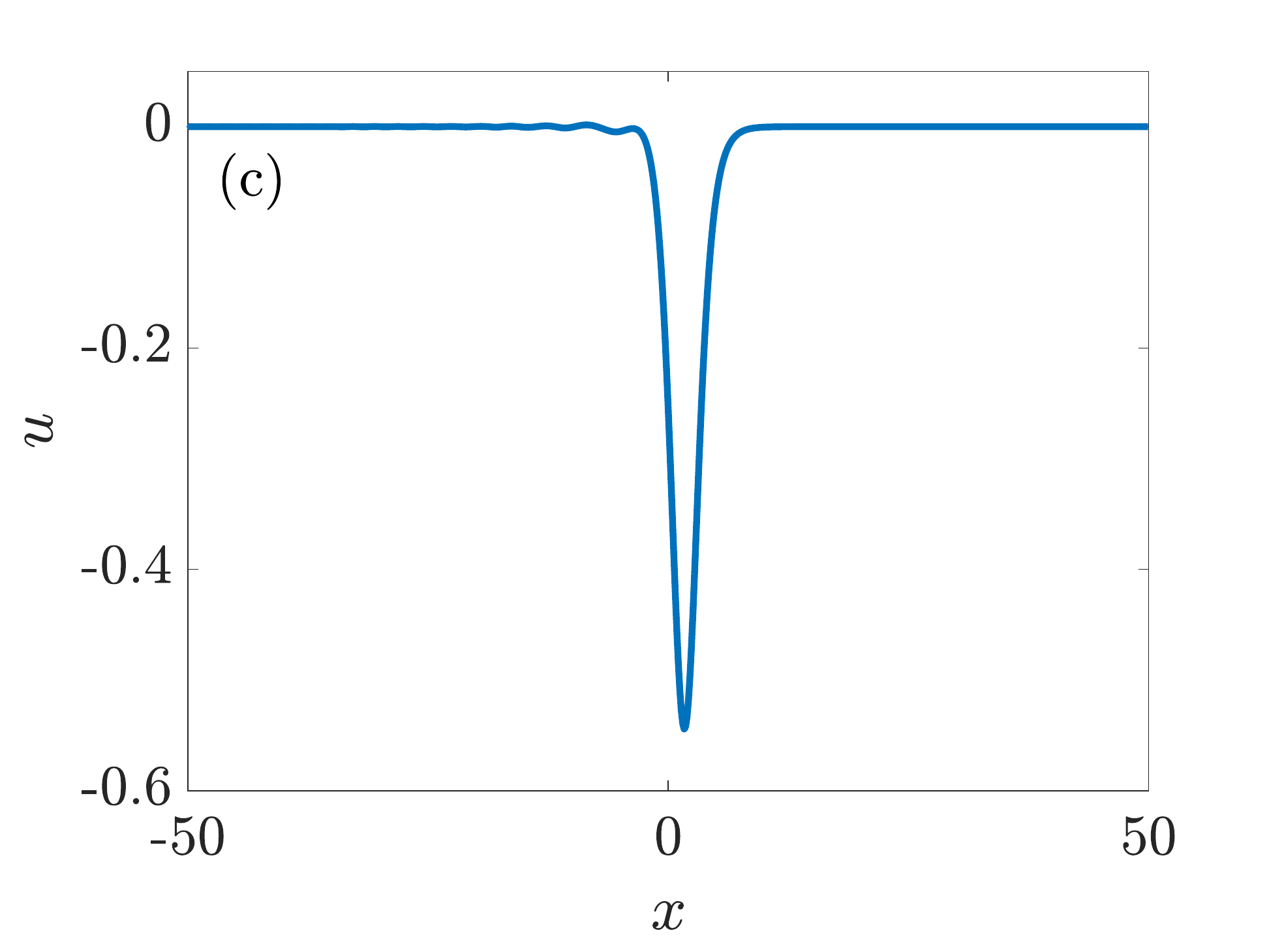}
\caption{%Top: Solution of \eqref{e.ux} for a fixed realisation of the noise with $\sigma=0.5$. Bottom left: Initial condition $u(x,t=0)$ with $w(0)=0.5$. Bottom right: Snapshot of the solution at $t=4.5$.
(a) Solution of \eqref{e.ux} for a fixed realisation of the noise with $\sigma=0.5$. (b) Initial condition $u(x,t=0)$ with $w(0)=0.5$ and $\kappa(0)=w^2(0)$. (c) Snapshot of the solution at $t=1.7$.}
\label{f.ux_waterfall}
\end{figure}

We again numerically solve the SPDE (\ref{e.ux}) using finite differences with periodic boundary conditions as described in Section~\ref{s.u}. We employ here a spatial discretisation of $\Delta x=0.15$ and an integration time step of $\Delta t=1\cdot 10^{-6}$. We show in Figure~\ref{f.ux_waterfall} the solution evolving from an initial soliton solution (\ref{e.uhat}) to an increasingly peaked solution, losing coherence by developing short-wave radiation which is amplified by the multiplicative noise involving the derivative of the solution. We remark that for finite discretisation $\Delta t$ and $\Delta x$ the scheme will develop numerical instabilities and the simulations develop machine-infinity at $t\approx 5$ (not shown).\\
	
%Simulations of the stochastically perturbed KdV equation (\ref{e.ux}) can be entirely understood from (\ref({e.v}). The perturbation causes an initial soliton to radiate waves. These waves then blow up according to the negative damping in (\ref{e.ux2}). Initially a soliton sheds a wave with wave length of its typical length scale $2/w_0$ with wave number $k_1=\pi w_0$. The energy of this linear wave will have grown by $5\%$ at some time $t_1=\log(1.05)/(\sigma^2k_1^2)\approx 0.08$. After $t>t_1$ additional numerical errors amplify and lead to a blow-up of the integration. This is seen in Figure~\ref{f.ux_waterfall}.\\

We now perform the collective coordinate approach outlined in Section~\ref{s.cc} for the collective coordinates ${\bf{c}}=\{\kappa,w,\phi\}$. We will see that the collective coordinate approach exhibits finite-time blow up with entirely deterministic dynamics for the amplitude and the inverse width and diffusive behaviour of the location of the coherent wave, consistent with the Galilean transformation leading to (\ref{e.ux2}). Furthermore, we show that we can estimate the time for which (\ref{e.ux}) can be used to describe coherent solitary waves.\\
	
We again seek the temporal evolution for the collective coordinates ${\bf{c}}=\{\kappa,w,\phi\}$ which we recall as
\begin{align*}
\ud \kappa&=a_\kappa\dt + \sigma_\kappa\dW,\\
\ud w&= a_{w}\dt+\sigma_w\dW,\\
\ud \phi&=a_\phi\dt+\sigma_\phi\dW.
\end{align*}
The projection of the residual as described in Section~\ref{s.cc} leads to the same  drift contributions (\ref{e.drift_u1})--(\ref{e.drift_u3})  of $\mathcal{O}(dt)$ as for $R(u)=u$. 
%\begin{align*}
%2 a_{\kappa}-\frac{\kappa }{w}a_w&=0,\\
%-\frac{\kappa }{w}a_{\kappa} + 2a_w&=0,\\
%a_{\phi}&=-\frac{4}{7}\left(5 w^2-12 \kappa \right),
%\end{align*}
The diffusion contributions of $\mathcal{O}(\sqrt{dt})$ are evaluated as 
\begin{align*}
\left(2\sigma_{\kappa} - \frac{\kappa }{w}\sigma_w \right)\dW&=0,\\
\left(-\frac{\kappa}{w}\sigma_{\kappa} + 2 \sigma_w \right) \dW&=0,\\
\sigma_{\phi}\dW&=-\sigma\ud B.
\end{align*}
Setting $\dW=\ud B$ we obtain 
\begin{align}
\ud\kappa&=\frac{2 \sigma ^2 \left(15+4 \pi ^2\right) }{5 \left(4 \pi ^2-15\right)}\kappa  w^2\dt,\\
\ud w&=\frac{24 \sigma ^2}{4 \pi ^2-15} w^3\dt,\\
\ud\phi&=\frac{4}{7}\left(12\kappa-5w^2\right)\dt-\sigma\dW.
\label{e.ux_cc_phieq}
\end{align}
As for the full SPDE (\ref{e.ux}) the amplitude and inverse width evolve deterministically and the noise only enters the position. The deterministic equations for $\kappa$ and $w$ can be solved analytically to obtain
\begin{align}
\kappa(t)&=\kappa_0(1-2aw_0^2\sigma^2t)^{-b},
\label{e.ux_cc1}
\\
w(t)&=w_0(1-2aw_0^2\sigma^2t)^{-\ha},
\label{e.ux_cc3}
\end{align}
with $a=24/(4\pi^2-15)\approx0.98$ and $b=(15+4\pi^2)/120\approx0.45$ and initial amplitude and inverse width $\kappa_0$ and $w_0$, respectively. This implies a blow up in finite time at $t=t_b=1/(2aw_0^2\sigma^2)$. The blow-up time, however, is far greater than the times in which the numerical scheme remains stable for the discretisation steps $\Delta x$ and $\Delta t$ used here.\\	
%Note that $t_b\gg t_1$; hence in numerical simulations of (\ref{e.ux}) we will always see the instability due to linear waves before we see the instability of the coherent solitary wave.\\
	
%We remark that the situation of deterministic shape parameters and the noise only entering the position of the solution, is linked to the transformation of  the stochastically perturbed KdV equation to the dissipative deterministic PDE (\ref{e.ux}) and mirrors the case of travelling waves in dissipative SPDEs discussed in \cite{CartwrightGottwald19}. The dissipative character of the PDE (\ref{e.ux2}) leads to strong contraction of the shape dynamics which controls the noise. The noise is however free to move along the neutrally stable translational symmetry group.\\

Note that the evolution of the shape parameters $\kappa$ and $w$ is deterministic and the noise only enters the location $\phi$ of the solution. This is linked to the transformation of the stochastically perturbed KdV equation (\ref{e.ux}) to the non-conservative deterministic PDE (\ref{e.ux2}) and mirrors the case of travelling waves in dissipative SPDEs discussed in \cite{CartwrightGottwald19}. The strong expansion of the PDE (\ref{e.ux2}) dominates the shape dynamics. The noise is however free to move along the neutrally stable translational symmetry group.\\
	
%\gaginline{The following paragraph might (hopefully) be used...not sure sure though, as the soliton constantly sheds waves ...? we will see.}
%In order to compare our result with simulations of the full stochastically perturbed KdV equation (\ref{e.ux}) we introduce a sponge layer
%\begin{align}
%\nu(x) = \nu_s(\tanh(a_s(x-x_s))+1)
%\end{align}	
%and solve, instead of (\ref{e.ux}) the equation 
%\begin{align}
%\ud u=(6uu_x-u_{xxx}+ \nu(x) u_{xx})\dt + \sigma u_x\ud B.
%\label{e.uxs}
%\end{align}
%This allows us to focus on the instability experienced by the actual solitary wave rather than on the instability experienced by the linear waves which are shed by the perturbed solitary wave.
%\gaginline{end of possible new paragraph}
	
\begin{figure}[htbp]
	\centering
	\includegraphics[width = 0.32\columnwidth]{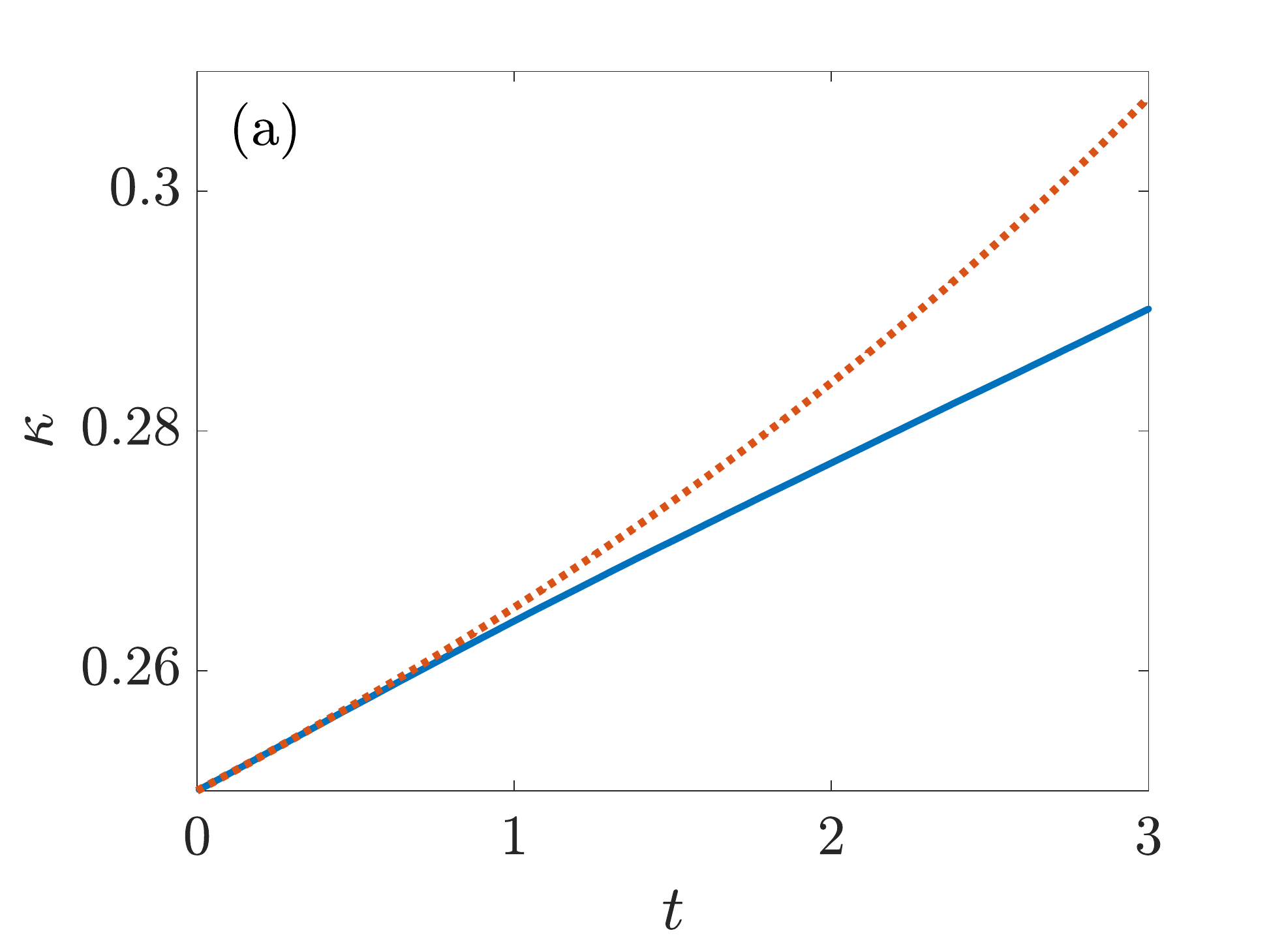}
	\includegraphics[width = 0.32\columnwidth]{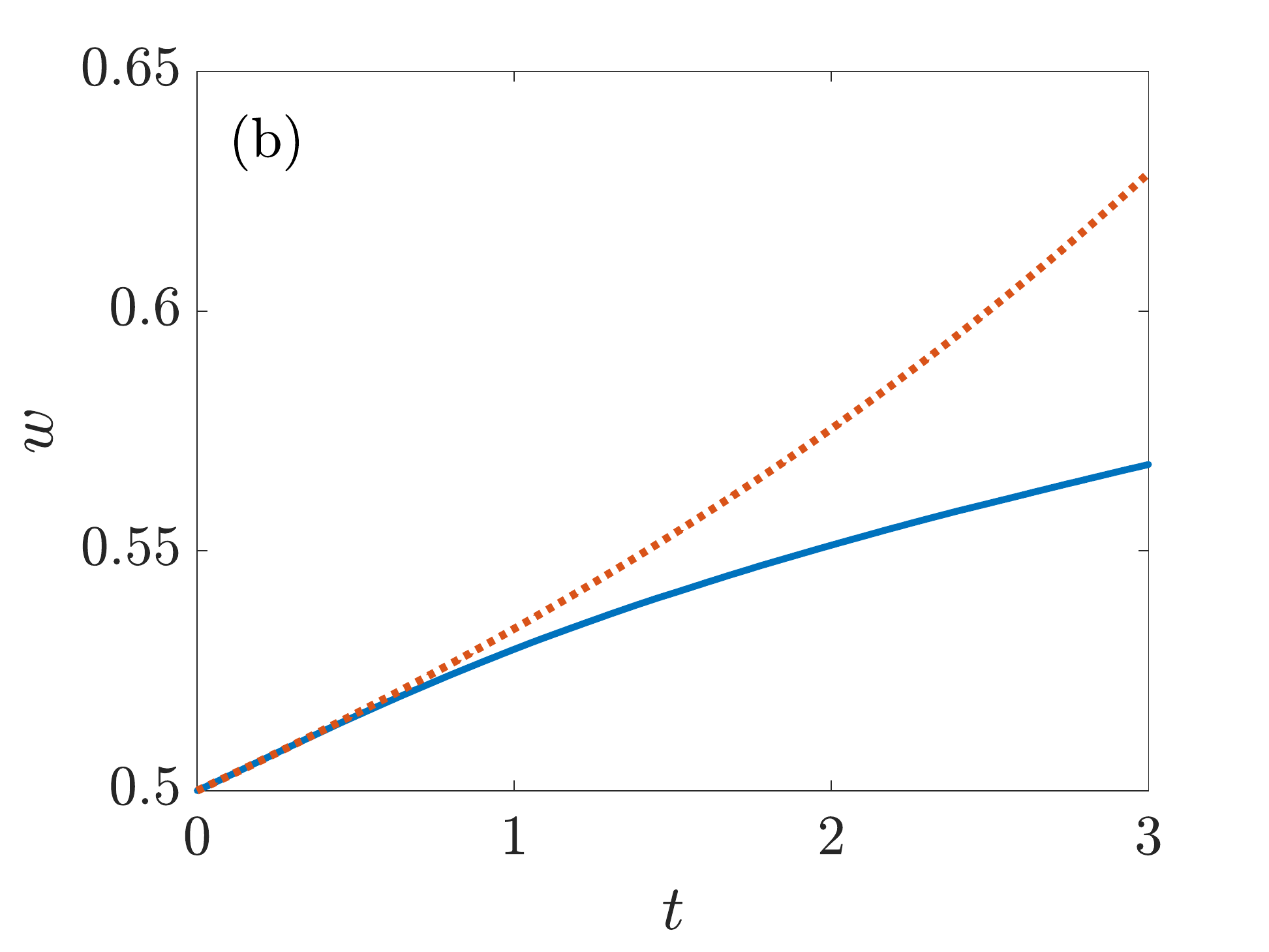}
	\includegraphics[width = 0.32\columnwidth]{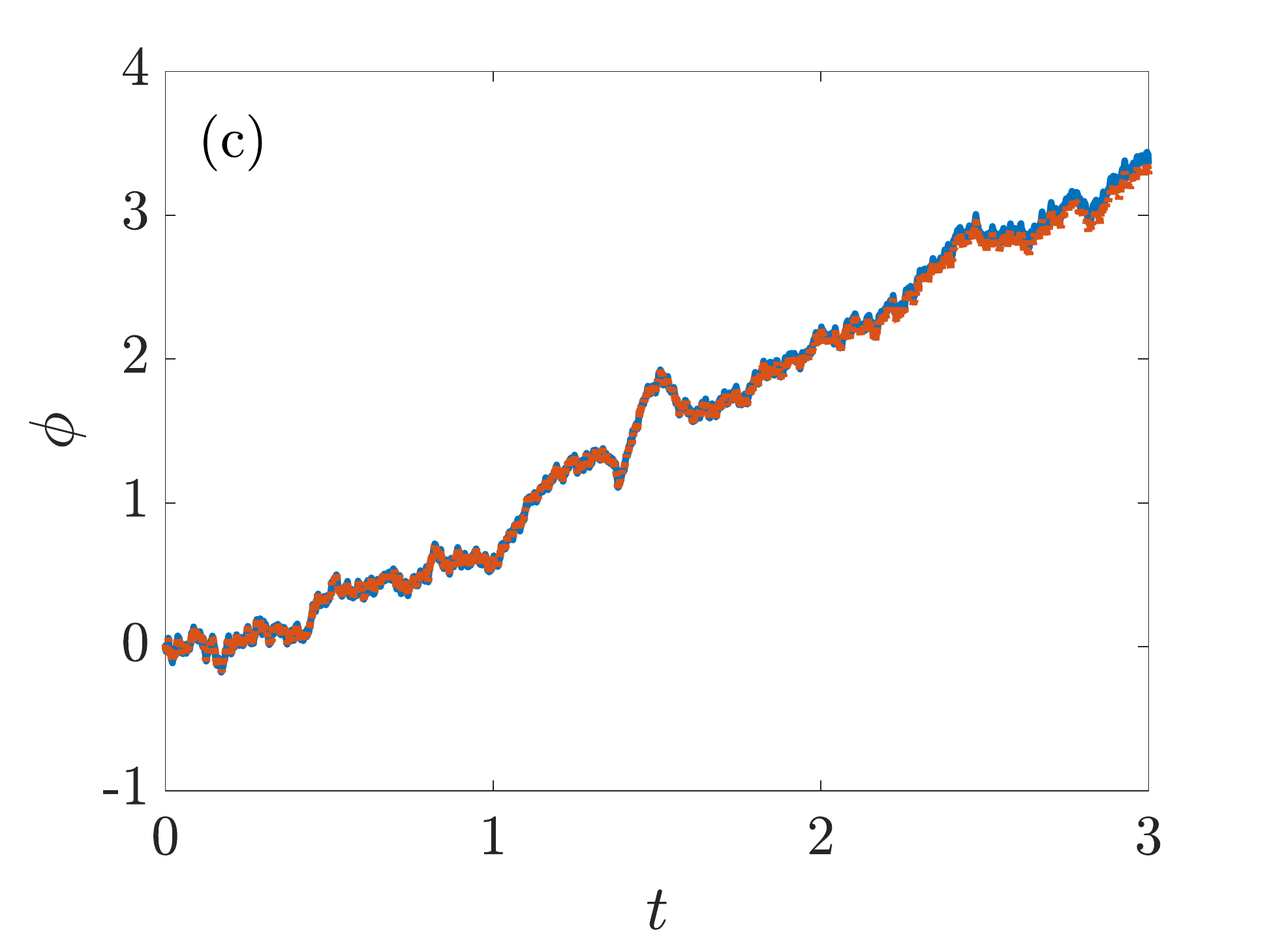}
\caption{Amplitude $\kappa$, inverse width $w$ and location $\phi$ of a solitary wave ansatz function (\ref{e.uhat}) for the stochastically perturbed KdV equation (\ref{e.ux}) as a function of time. Continuous lines (online blue) are obtained from a direct simulation of (\ref{e.ux}). The dotted lines (online red) are the results from the collective coordinate approach (\ref{e.ux_cc1})--(\ref{e.ux_cc3}). %The vertical bars are at $t=???$. 
%for which the absolute error of $w$ is $2\%$. 
Parameters as in Fig~\ref{f.ux_waterfall}.}
%\caption{Time series of $\kappa$, $w$ and $\phi$ for fixed noise strength $\sigma=0.5$ and initial inverse width $w_0=0.5$. The collective coordinates solutions are given in red, while the fits to the full solution obtained by numerical integration of the SPDE are in blue. Top left: Plot of $\kappa$ over time. Note that the fit to the true solution shown is the average over 100 realisations of the noise to account for the instability of the solver??? Top right: Plot of $\kappa$ over time. Note that the fit to the true solution shown is the average over 100 realisations of the noise to account for the instability of the solver??? Bottom: Plot of $\kappa$ over time for a particular realisation of noise. Note that the two are hard to distinguish from each other in the main plot due to their similarity to each other. The inset is a close up of the end of the time series to allow the differences between the collective coordinates solution and fit to the full solution to be observed more easily. }
\label{f.ux_cc_t}
\end{figure}

Figure~\ref{f.ux_cc_t} shows a comparison of the collective coordinate approach (\ref{e.ux_cc1})--(\ref{e.ux_cc3}) with a numerical simulation for the stochastically perturbed KdV equation (\ref{e.ux}) with $\sigma=0.5$ for one realisation of the noise. The deterministic behaviour of the amplitude $\kappa$ and the inverse width $w$ is clearly contrasted to the diffusive dynamics of the location $\phi$. The values the collective coordinates corresponding to the solution of the direct numerical simulation of the SPDE are again obtained via a nonlinear least square fit to the ansatz solution (\ref{e.uhat}). The collective coordinate approach yields a remarkably good approximation for some time until it deteriorates after $t\approx1.7$. The location $\phi(t)$ is particularly well described by the collective coordinate equation (\ref{e.ux_cc3}) and tracks the location of the solitary wave in the full SPDE for much longer times than achieved by the amplitude and inverse width. The reason for this is that the dynamics of $\phi$ (\ref{e.ux_cc_phieq}) is noise-dominated. % until $t=0.015$ before the drift starts dominating.

\begin{figure}[htbp]
	\centering
	\includegraphics[width = 0.7\columnwidth]{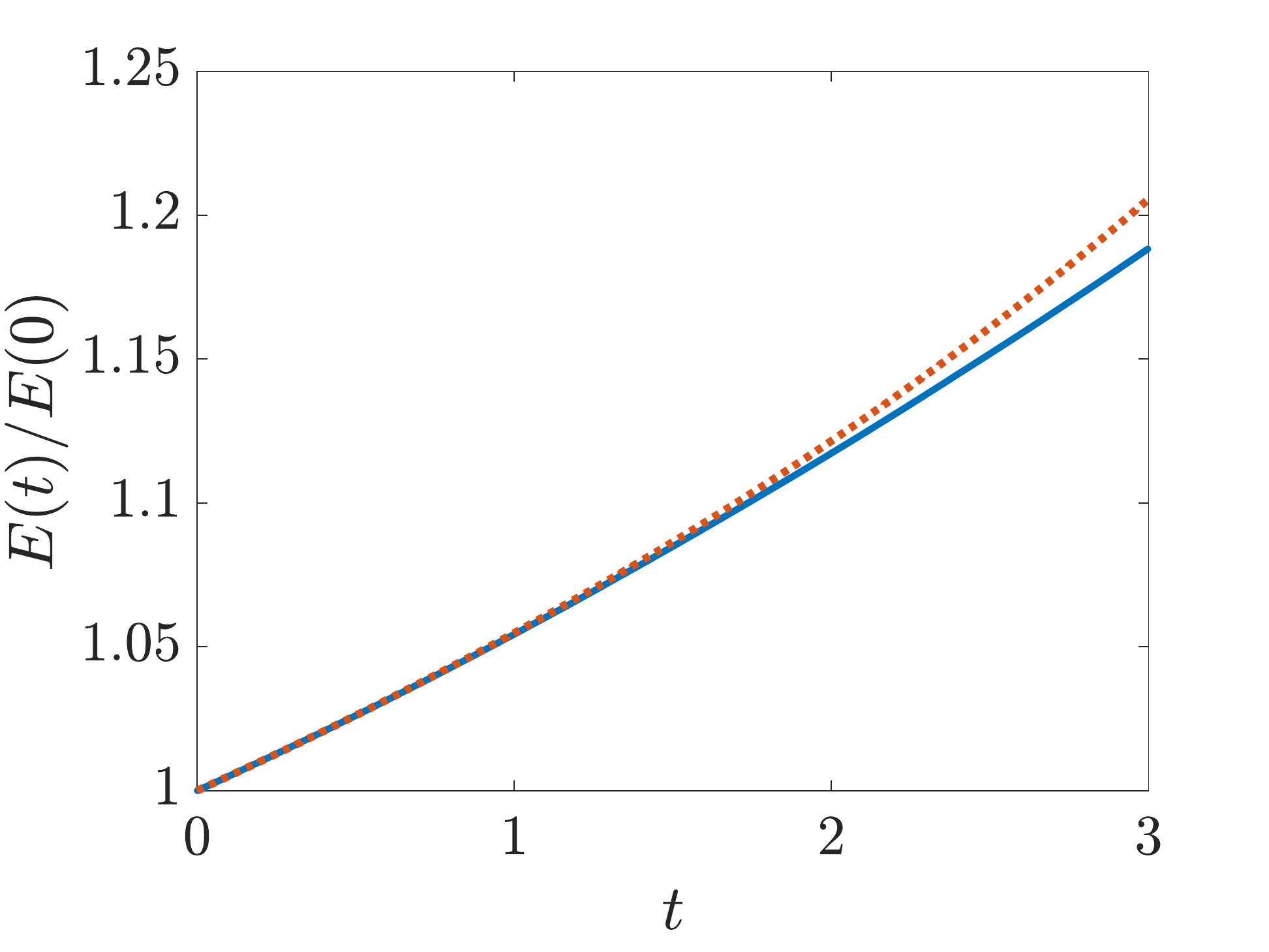}
\caption{Normalized Energy $E(t)/E(0)$ calculated from a numerical simulation of the stochastically perturbed KdV equation (\ref{e.ux}) (continuous lines, online blue) and calculated using collective coordinates (dotted lines, online red). Parameters as in Fig~\ref{f.ux_waterfall}.}
\label{f.ux_energysq_t}
\end{figure}

Figure~\ref{f.ux_energysq_t} shows that the energy $E$ is very well tracked by the corresponding energy of the collective coordinate ansatz 
\begin{align*}
E_{\rm{cc}}(t)&=\int \hat u^2(x;\kappa,w,\phi)\, dx % =  \frac{16}{3}\frac{\kappa^2}{w} 
\\
&= \frac{16\kappa_0^2}{3w_0} \, \left(1-2aw_0^2\sigma^2 t\right)^\frac{4b-1}{2},
\end{align*}
with $\kappa_0=\kappa(0)$ and $w_0=w(0)$. Note that the energy tracks the energy of the SPDE for longer times than the amplitude and inverse width individually. 

We estimate the time for which coherence is ensured in the sense that the solution of (\ref{e.ux}) can be well approximated by a coherent solitary wave of the form (\ref{e.ux}), by estimating the time for which the energy, as calculated by the collective coordinates, has grown to a value of $10\%$ of its initial value. We hence define the time of coherence $\tau_c$ as $E_{\rm{cc}}(\tau_c)=1.1 E_{\rm{cc}}(0)$, leading to
\begin{align}
%\tau_c =\frac{1-1.1^{2/(4b-1)}}{2aw_0^2\sigma^2}=\frac{(4\pi^2-15)(1-1.1^{60/(15-4\pi^2)})}{48w_0^2\sigma^2}\approx\frac{0.106}{w_0^2\sigma^2}.
\tau_c =\frac{1-1.1^{\tfrac{2}{1-4b}}}{2a w_0^2\sigma^2}
\approx\frac{0.106}{w_0^2\sigma^2}.
\label{e.ux_tauc}
\end{align}
Figure~\ref{f.ux_waterfall} shows the solution at $t=\tau_c=1.7$ for $\sigma=0.5$ and $w_0=0.5$. For $t>4$ the exponential growth of the high wavenumbers generated by the steepening of the wave will have amplified to destroy the solution.

%\begin{figure}[htbp]
%	\centering
%	\includegraphics[width = 0.6\columnwidth]{ux_l2err.pdf}
%	\caption{$L_2$-difference between the simulated solution to the full SPDE \eqref{e.ux} and our collective coordinates solution to (\ref{e.ux_cc1})--(\ref{e.ux_cc3}). 
%		Parameters as in Fig~\ref{f.ux_waterfall}.}
%	\label{f.ux_tauc}
%\end{figure}
%\gaginline{Need to define $L_2$-difference.}
%%%%%%%%%%%%%%%%%%%%%%%%%%%%%%%%%%%%%%%%%%%%%%%%%%%%%%%%%%%%%%%%%
	
\section{Conclusion} 
\label{s.discussion}

We presented a collective coordinate framework to study the dynamics of solitary waves in stochastically perturbed Korteweg-de Vries equations. Different to previous collective coordinate approaches for the KdV equation which were developed in the deterministic context and had a hard-coded constraint between the amplitude of the traveling solitary wave and its inverse half width, we treat them as independent parameters. This was shown to deal better with larger perturbations which occur in stochastically driven KdV equations.

We studied homogeneous additive noise as well multiplicative noise. Our collective coordinates was able to recover the well-known analytical solution for the additive noise \citep{Wadati83}. The case of multiplicative noise with $R(u)=u$ leads to the solitary wave loosing coherence via radiation. This effect is typically not described by collective coordinate approaches which focus on the coherent part. The case of multiplicative noise with $R(u)=u_x$ leads to an  ill-posed SPDE. It is nevertheless used to model waves with fluctuating velocities in situations where the noise is confined to sufficiently small spatial domains. In this case, coherence is lost by increasing energy and by small-scale features getting amplified.

We used the reduced description of the collective coordinates to design diagnostics which allowed us to quantify the time of coherence of a solitary wave experiencing stochastic perturbations.  The diagnostics is dependent on the way coherence is lost. For the case $R(u)=u$ we monitored coherence by quantifying the systematic noise-driven deviation from the deterministic location of the solitary wave. For the ill-posed case $R(u)=u_x$ we monitored the increase of the energy as estimated by the collective coordinates. In both cases the estimate of the coherence time can be achieved by only using information of the reduced collective coordinate dynamics.

From a modelling perspective our collective coordinate framework can be used to determine the time of validity. If a modeller is interested in studying the effect of random perturbations on a coherent wave, then our decoherence time can be used to estimate the time-scale for which models invoking coherent solutions are valid.

%%%%%%%%%%%%%%%%%%%%%%%%%%%%%%%%%%%%%%%%%%%%%%%%%%%%%%%%%%%%%%%%%

\section*{Acknowledgments}
We acknowledge funding by the Australian Research Council, Grant No.~DP180101991.
	
%%%%%%%%%%%%%%%%%%%%%%%%%%%%%%%%%%%%%%%%%%%%%%%%%%%%%%%%%%%%%%%%%
	
\appendix
	
\section{Explicit formulae for the collective coordinate projections} 
\label{a.formulae}
We list here several integrals which appear in the evaluations of the projection when using the ansatz function (\ref{e.uhat}) for the solitary wave, which we recall here 
\begin{align*}
\hat u(x,t) =-2\kappa(t) \sechsq{w(t)(x-\phi(t))}+\beta(t).
\end{align*}
Using $\uh_x=-\pdif{\uh}{\phi}$, $\uh_{\kappa}=\frac{\uh}{\kappa}$, $\uh_{\kappa w}=\frac{1}{\kappa}\pdif{\uh}{w}$, $\uh_{\kappa\phi}=-\frac{1}{\kappa}\pdif{\uh}{x}$, $\uh_{xxx}=-\pdif{^3\uh}{\phi^3}$, and  $\uh_{\beta}=1$, we evaluate (omitting the hats for ease of exposition)
\begin{align*}
&\langle u^2\rangle = \frac{16 \kappa ^2}{3 w},
\quad
\langle u\pdif{u}{w}\rangle = -\frac{8 \kappa ^2}{3 w^2},
\quad
\langle \left(\pdif{u}{w}\right)^2\rangle=\frac{16 \pi ^2 \kappa ^2}{45 w^3},
\quad
\langle \left(\pdif{u}{x}\right)^2\rangle=\frac{64 \kappa ^2 w}{15},
\\
&\langle u\rangle=-\frac{4 \kappa }{w},
\quad
\langle\pdif{u}{w}\rangle=\frac{4 \kappa }{w^2},
\quad
\langle\pdif{u}{x}\pdif{^2u}{w\partial x}\rangle=\frac{32 \kappa ^2}{15},
\quad
\langle \pdif{u}{x}\pdif{^3u}{x^3}\rangle=-\frac{256}{21} \kappa ^2 w^3,\\
&\langle u \pdif{u}{x}\rangle =\langle \pdif{u}{w} \pdif{u}{x}\rangle=\langle u \pdif{^2u}{w\partial x}\rangle=\langle \pdif{u}{w} \pdif{^2u}{w\partial x}\rangle =0, \\
& \langle u\pdif{^3u}{x^3}\rangle=\langle \pdif{u}{w}\pdif{^3u}{x^3}\rangle=\langle u^2 \pdif{u}{x}\rangle=\langle u\pdif{u}{x}\pdif{u}{w}\rangle=\langle \pdif{u}{x}\rangle=0.\\
\end{align*}

We further list integrals that appear in the calculations for the Lagrangian variational framework outlined in Appendix~\ref{a.lagrange}. Here we have $\psi_x=u$, i.e. 
\[\psi(x,t)=-\frac{2\kappa}{w}\tanh\left(w(x-\phi)\right). \]
Then $\psi_{\kappa}=\frac{\psi}{\kappa}$, $\psi_{\phi}=-\pdif{\psi}{x}=-u$, $\psi_{\kappa w}=\frac{1}{\kappa}\pdif{\psi}{w}$, $\psi_{\kappa\phi}=-\frac{1}{\kappa}\pdif{\psi}{x}=-\frac{u}{\kappa}$ and $\psi_{w\phi}=-\pdif{^2\psi}{w\partial x}=-u_w$, and we evaluate
\begin{align*}
&\langle\left(\pdif{\psi}{x}\right)^3\rangle=\langle u^3\rangle = -\frac{128 \kappa ^3}{15 w},
\quad
\langle\left(\pdif{^2\psi}{x^2}\right)^2\rangle=\langle u_x^2\rangle=\frac{64 \kappa ^2 w}{15},\\
&\langle\psi\pdif{\psi}{x}\rangle=\langle\pdif{\psi}{x}\pdif{\psi}{w}\rangle=\langle\pdif{\psi}{x}\pdif{^2\psi}{w^2}\rangle=0.
\end{align*}

%%%%%%%%%%%%%%%%%%%%%%%%%%%%%%%%%%%%%%%%%%%%%%%%%%%%%%%%%%%%%%%%%

\section{Collective coordinate approach within a Lagrangian variational framework}
\label{a.lagrange}
We provide here a stochastic version of the well-known variational collective coordinate approach within a Lagrangian formulation for deterministic perturbations \citep{Whitham,AndersonEtAl88,BassEtAl88,KivsharMalomed89,ScottBook}, which to the best of our knowledge has not been presented in the literature. We then present a numerical illustration of a deterministically perturbed KdV equation illustrating the differences between our approach, based from the point of view of Galerkin approximations, and the Lagrangian approach.\\
	
Consider perturbations $P(u,x,t)$ of the KdV equation in the form
\begin{align}
u_t - 6uu_x + u_{xxx} = P(u,x,t).
\label{e.pkdv}
\end{align}
Here $P(u,x,t)$ may be a deterministic or stochastic perturbation, with the obvious interpretation. The integrable KdV equation with $P\equiv 0$ is variational with Lagrangian density
\begin{align}
\mathcal{L} = \frac{1}{2}\psi_t\psi_x -\psi_x^3-\frac{1}{2}\psi_{xx}^2
\end{align}
with $\psi_x = u$. The solution for the KdV equation is given by (\ref{e.soliton}) which we recall here
\begin{align}
u(x,t)=-2\kappa \sechsq{w(x+\phi)},
\label{e.solitonA}
\end{align}
with amplitude $\kappa=w^2$ and location $\phi=4w^2 t$. Assuming that the collective coordinates ${\bf{c}}=\{\kappa,w,\phi\}$ are time-dependent, upon substitution of the ansatz solution (\ref{e.solitonA}) the Lagrangian can be evaluated as
\begin{align}
L=\int\mathcal{L}dx = -\frac{8}{3}[ 
\frac{\kappa^2}{w}\dot \phi 
+ \frac{\kappa}{w}\dot \kappa \dot \phi 
- \frac{1}{2}\frac{\kappa^2}{w^2}\dot w\dot \phi 
- \frac{48}{15}\frac{\kappa^3}{w}
+\frac{12}{15}\kappa^2 w
].
\label{e.L}
\end{align}
The relevant integrals used to obtain (\ref{e.L}) are listed in Appendix~\ref{a.formulae}. Note that we included quadratic terms of time-derivatives to adhere to It\^o calculus for eventual stochastic perturbations; for deterministic perturbations these quadratic terms are to be discarded. For simplicity we do not include here the constant background term $\beta$ (cf (\ref{e.uhat})) as a collective coordinate. 
	
The Euler-Lagrange equations for a collective coordinate $c_j$ is calculated as
\begin{align*}
\frac{d}{dt}\left(\frac{\partial L}{\partial {\dot c}_j} \right) -\frac{\partial L}{\partial c_j} 
& = 
\int \left(
\frac{\partial \mathcal{L}}{\partial \psi}
-\frac{\partial}{\partial t}  \frac{\partial \mathcal{L}}{\partial \psi_t}
-\frac{\partial}{\partial x}  \frac{\partial \mathcal{L}}{\partial \psi_x}
+ \frac{\partial^2}{\partial x^2}  \frac{\partial \mathcal{L}}{\partial \psi_{xx}}
\right) \frac{\partial \psi}{\partial c_j}\, dx\\
& = 
\int P(u,x,t)\frac{\partial \psi}{\partial c_j}\, dx.
%\int\left( \frac{\partial \mathcal{L}}{\partial \psi}\frac{\partial \psi}{\partial  \right) 
\end{align*}
For the Lagrangian (\ref{e.L}) the Euler-Lagrange equations become
\begin{align}
2\kappa\dot\kappa - \frac{\kappa^2}{w}\dot w + 2 {\dot \kappa}^2+2\frac{\kappa^2}{w^2}{\dot w}^2 -4\frac{\kappa}{w}\dot\kappa\dot w-\frac{1}{2}\frac{\kappa^2}{w}\ddot w &=\langle R(u) \frac{\partial \psi}{\partial \kappa}\rangle 
\label{e.lagc1}
\\
\dot \phi -\frac{4}{5}(6\kappa-w^2)-\frac{1}{2}\ddot\phi&=\langle R(u) \frac{\partial \psi}{\partial w}\rangle 
\label{e.lagc2}
\\
\dot \phi -\frac{4}{5}(4\kappa+w^2)-\frac{1}{2}\ddot \phi&=\langle R(u) \frac{\partial \psi}{\partial \phi}\rangle .
\label{e.lagc3}
\end{align}
Again terms containing two time derivatives originate from the application of It\^o calculus, and have to be discarded for deterministic perturbations. 
	
One can now proceed again by assuming
\begin{align*}
\ud\kappa&=a_{\kappa}\dt+\sigma_{\kappa}\dW,\\
\ud w&= a_w\dt+\sigma_w\dW,\\
\ud\phi&=a_{\phi}\dt+\sigma_{\phi}\dW,
\end{align*}
to determine the drift and diffusion terms for each of the collective coordinates as done in Section~\ref{s.cc}. The second time-derivatives in  (\ref{e.lagc1})--(\ref{e.lagc3}) contain contributions $\dW^2=\dt$, which we write here in differential form as
\begin{align*}
d(dw) &= \left( 
\frac{\partial \sigma_w}{\partial \kappa} \sigma_\kappa 
+\frac{\partial \sigma_w}{\partial w}\sigma_w
+ \frac{\partial \sigma_w}{\partial \phi} \sigma_\phi
\right) \dt + o(\dt) ,
\\
d(d\phi) &= \left( 
\frac{\partial \sigma_\phi}{\partial \kappa} \sigma_\kappa 
+\frac{\partial \sigma\phi}{\partial w}\sigma_w
+ \frac{\partial \sigma_\phi}{\partial \phi} \sigma_\phi
\right) \dt  + o(\dt).
\end{align*}
	
The resulting equations are different to those derived in Section~\ref{s.cc} which did not make explicit use of the variational structure of the KdV equation. For example, note that from (\ref{e.lagc2}) and (\ref{e.lagc3}) we conclude that $\kappa=\kappa(w,\phi)$ is algebraically constrained and hence the dynamics evolves in a two-dimensional subspace. This is in stark contrast to our framework where the collective coordinates evolve independently in $\R^3$. To  illustrate further the differences we consider now the deterministic perturbation of a linearly damped KdV equation with $P(u,x,t)=-\nu u=\nu \psi_x$. Since $\langle P(u,x,t) \psi_\kappa\rangle =  \langle P(u,x,t) \psi_w\rangle$ and $\langle P(u,x,t) \psi_\phi\rangle=\tfrac{16}{3}\nu\tfrac{\kappa^2}{w}$, the evolution equations for the collective coordinates (\ref{e.lagc1})--(\ref{e.lagc3}) become, upon discarding the terms involving two time-derivatives, 
\begin{align}
\dot \kappa &= -\frac{4}{3}\nu \kappa
\label{e.lcc1}
\\
\dot \phi &= 4 w^2 
\label{e.lcc2}
\end{align}
with the unperturbed algebraic solitary wave constraint $\kappa=w^2$. 
	
Our collective coordinate approach on the other hand yields, evaluating the integrals in Section~\ref{s.cc},
\begin{align}
\dot \kappa &= -\nu\kappa
\label{e.occ1}
\\
\dot w &=0
\label{e.occ2}
\\
\dot \phi &= \frac{4}{7}\left(12\kappa-5w^2\right) .
\label{e.occ3}
\end{align}
%Hence the variational framework forces a solitonic shape with $\kappa=w^2$. 
	
\begin{figure}[htbp]
	\centering
	\includegraphics[width = 0.32\columnwidth]{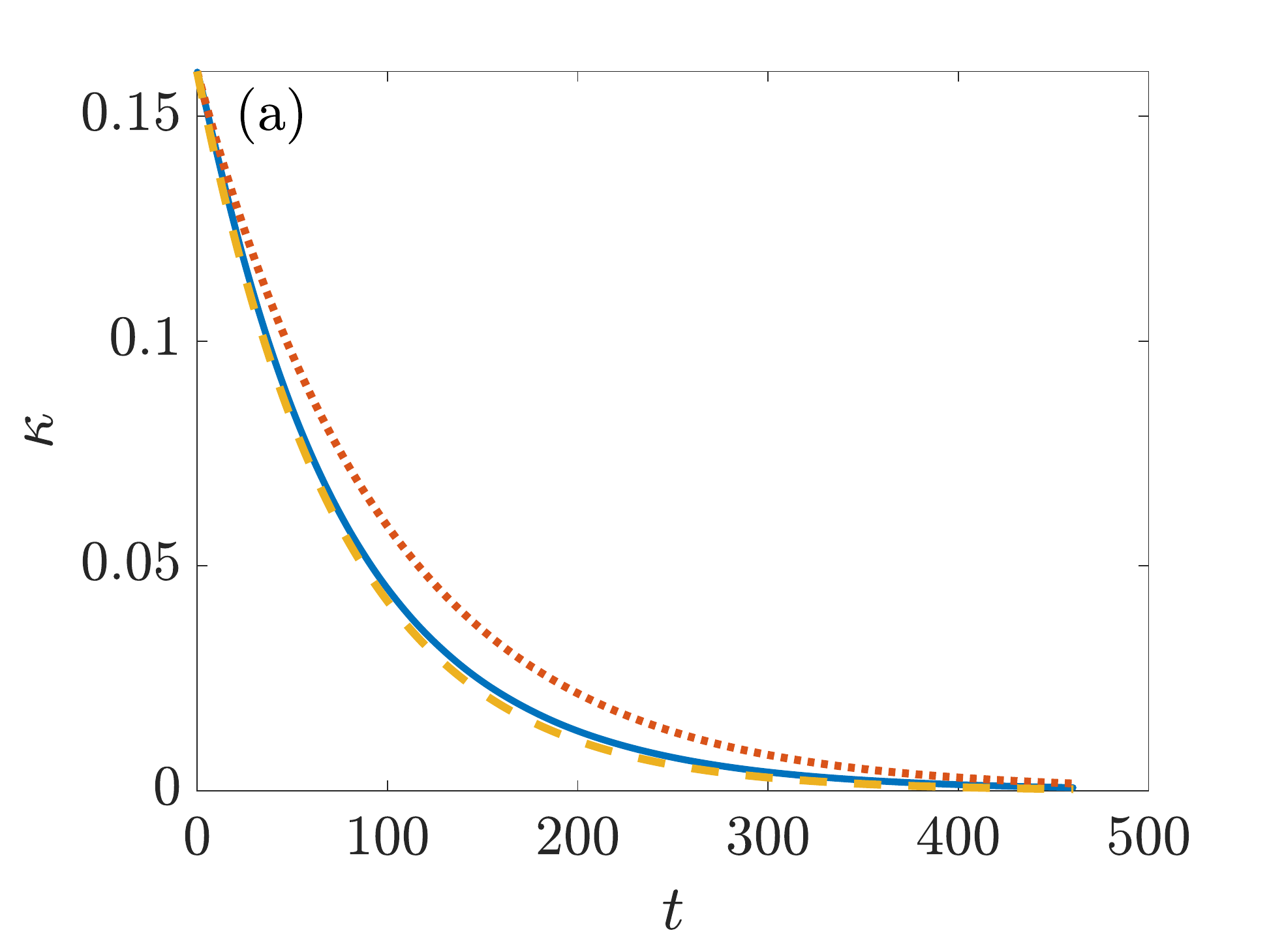}
	\includegraphics[width = 0.32\columnwidth]{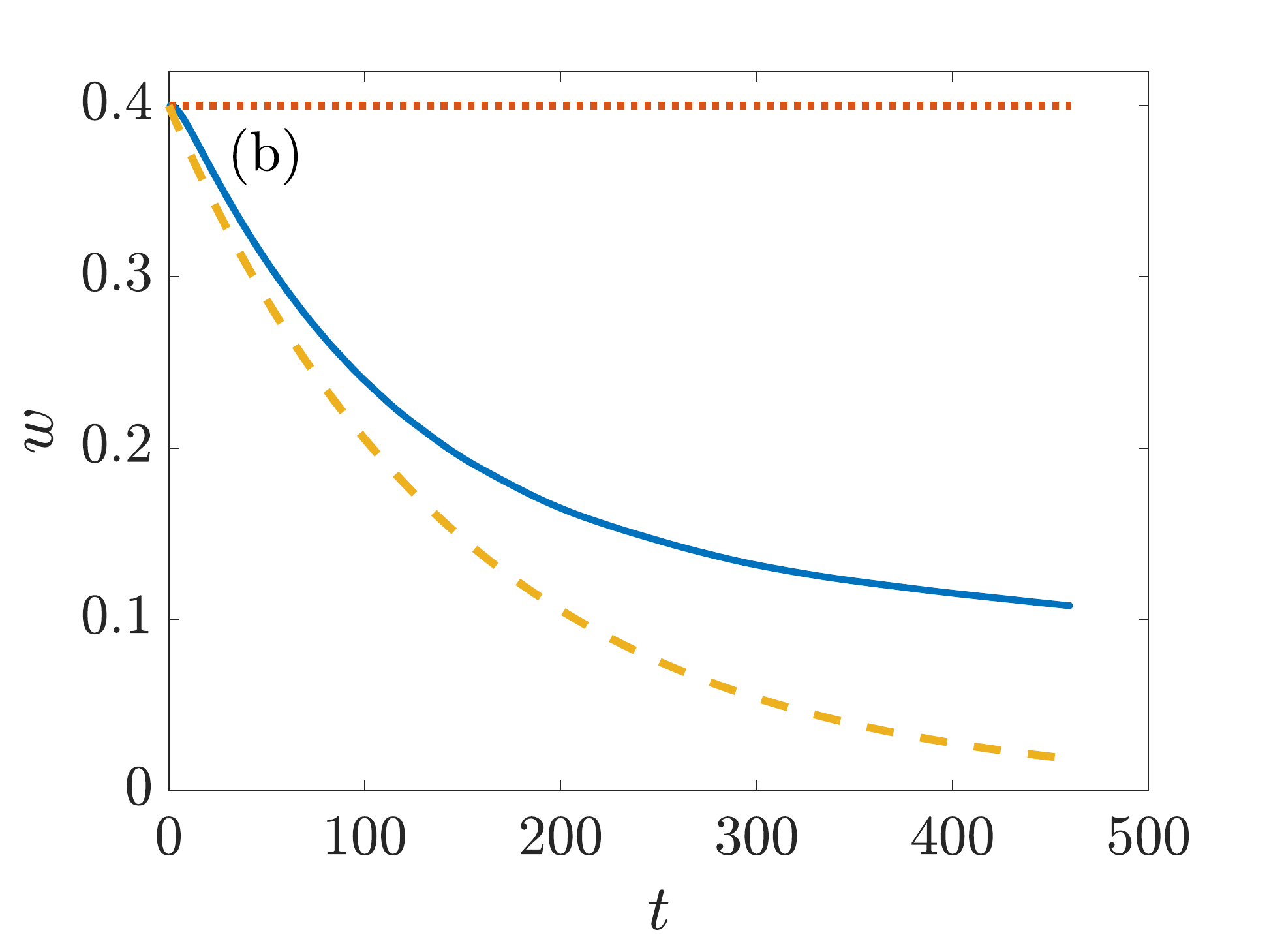}
	\includegraphics[width = 0.32\columnwidth]{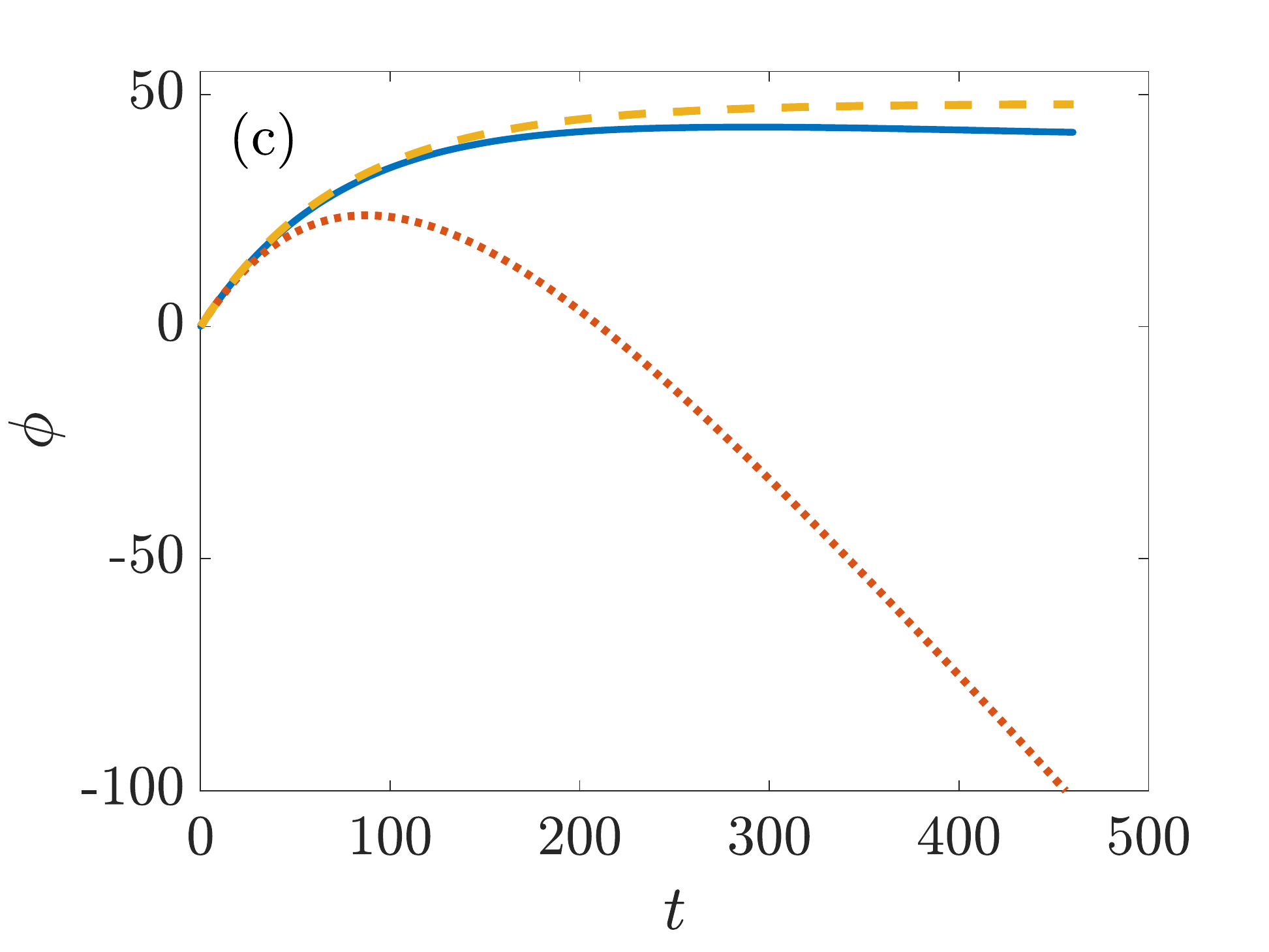}
\caption{Plot of the collective coordinates for deterministic damping $P(u,x,t)=-\nu u$ with $\nu=0.01$ as determined by our collective coordinate approach (\ref{e.occ1})--(\ref{e.occ3}) (dotted lines, online red), the Lagrangian collective coordinate approach (\ref{e.lcc1})--(\ref{e.lcc2}) (dashed lines, online yellow). The continuous line (online blue) depicts results of a simulation of the full perturbed KdV equation (\ref{e.pkdv}).}
\label{f.l_nu0p01}
\end{figure}

\begin{figure}[htbp]
	\centering
	\includegraphics[width = 0.32\columnwidth]{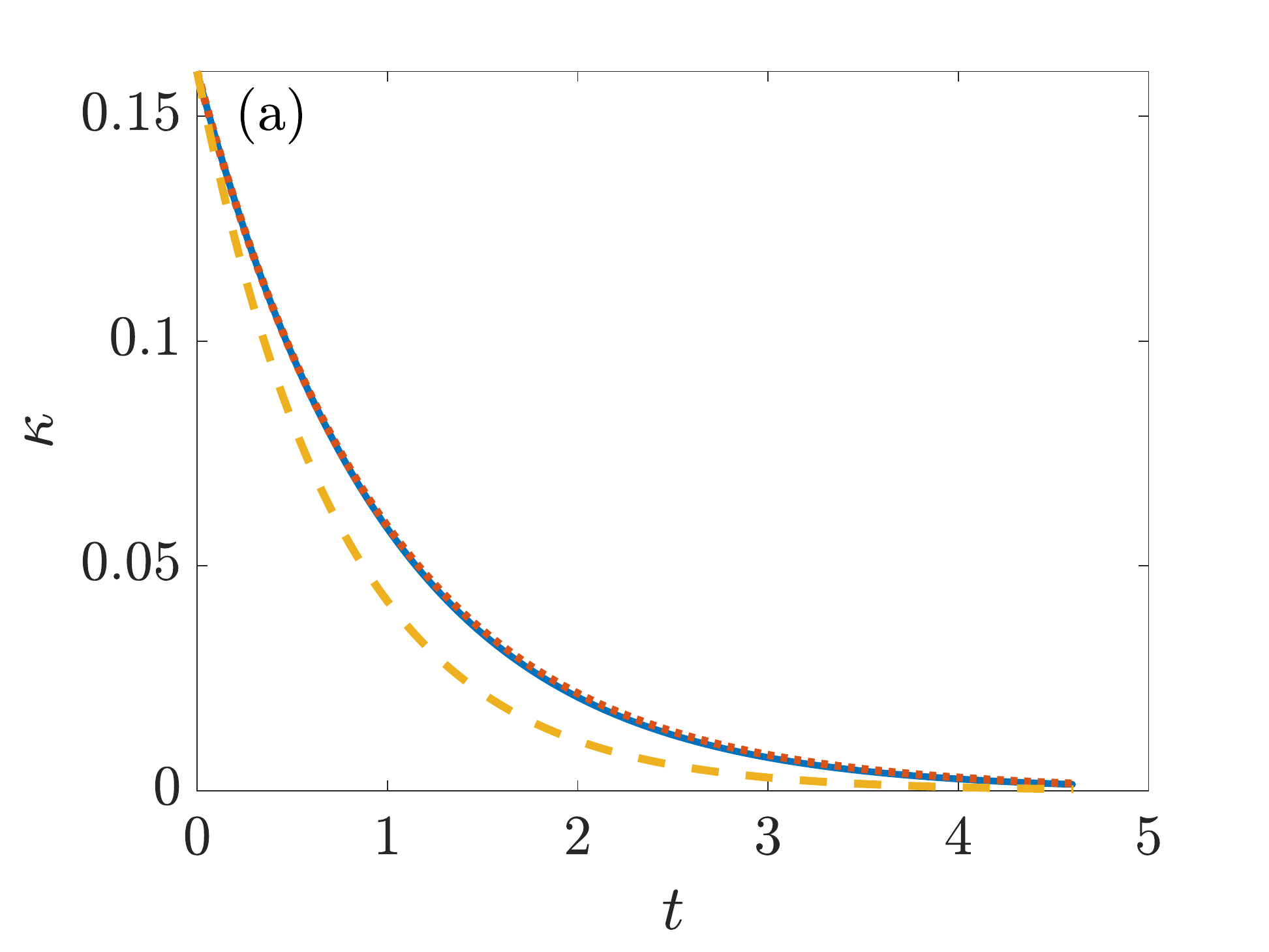}
	\includegraphics[width = 0.32\columnwidth]{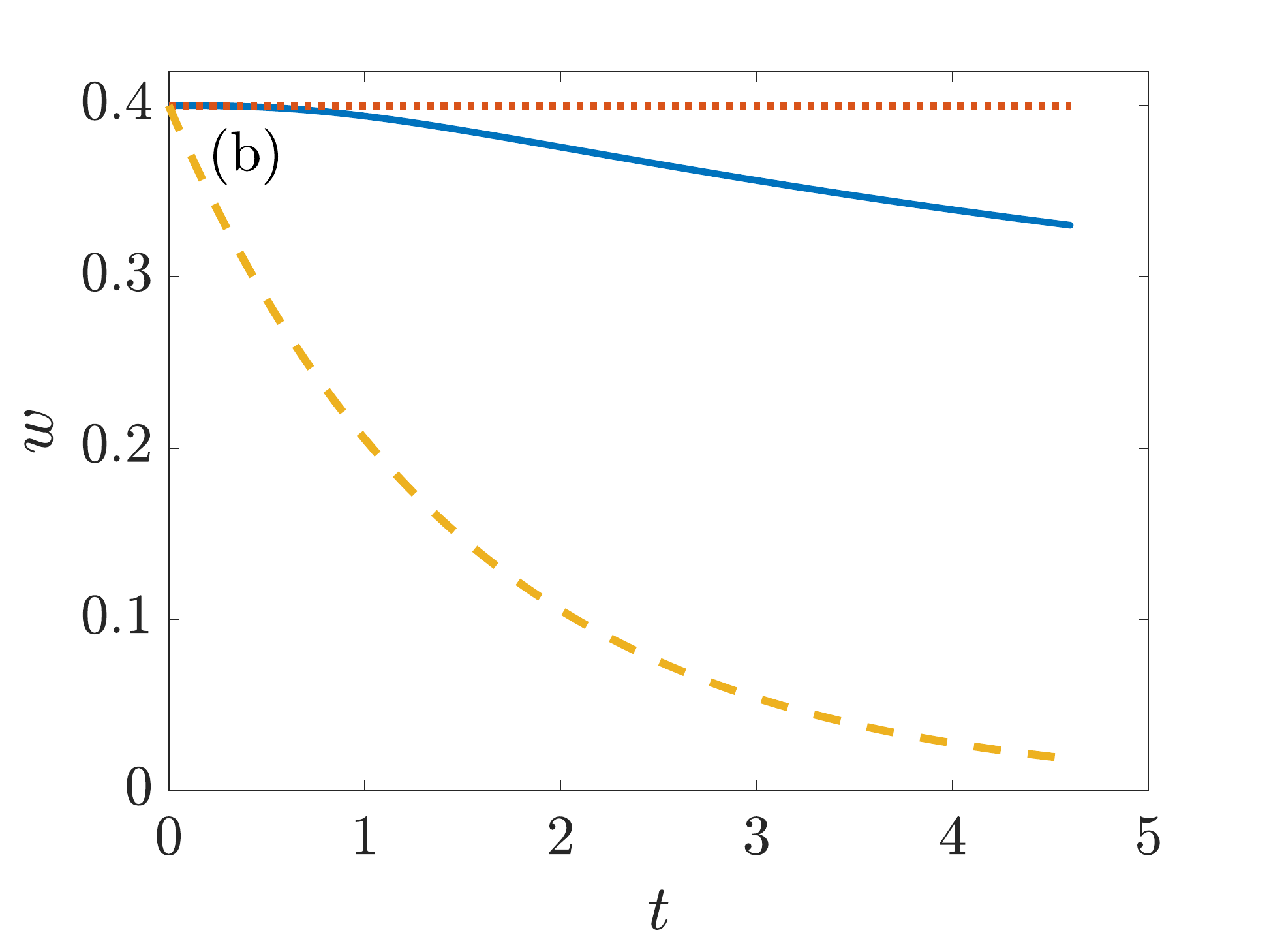}
	\includegraphics[width = 0.32\columnwidth]{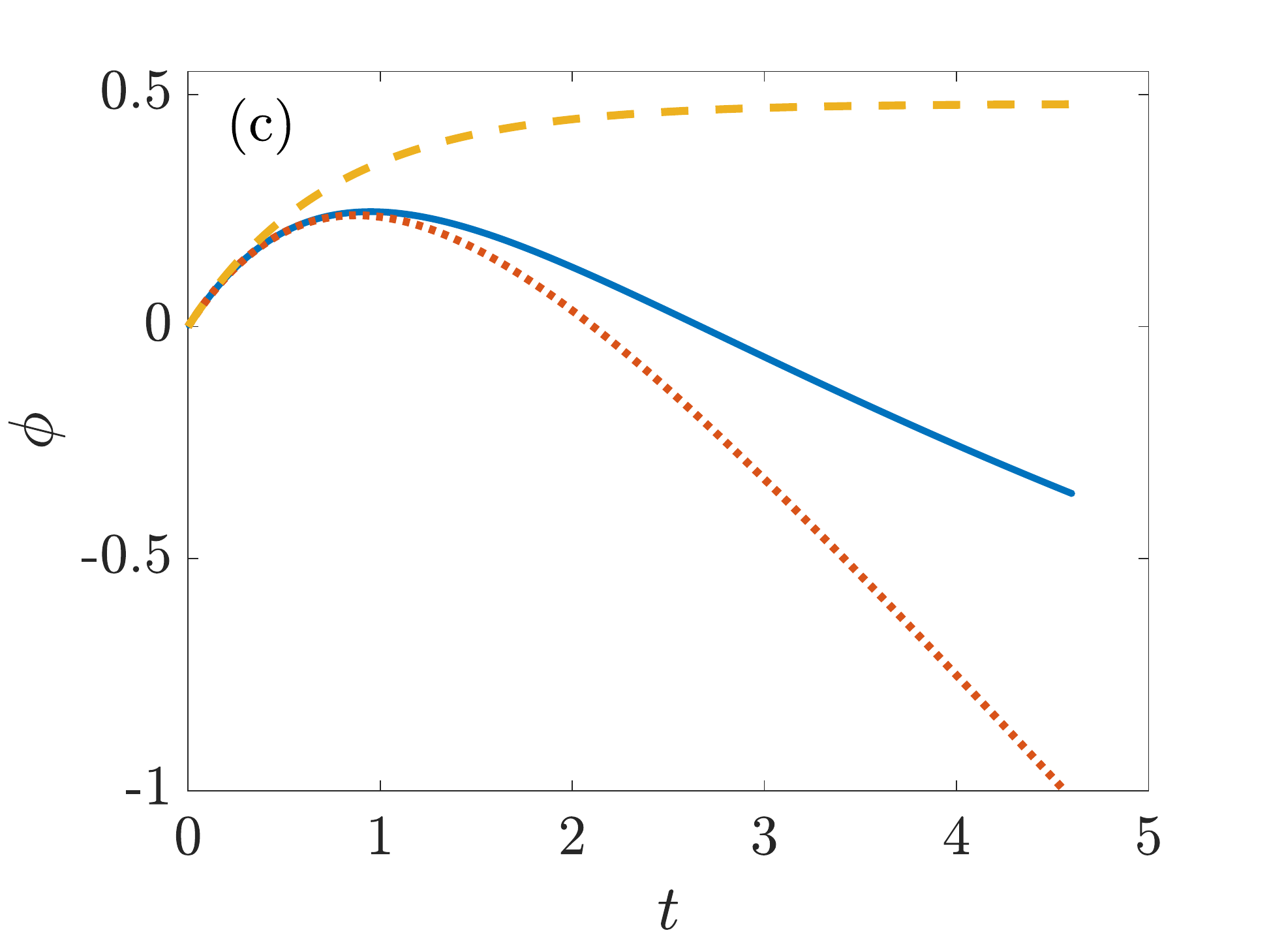}
\caption{Plot of the collective coordinates for deterministic damping $P(u,x,t)=-\nu u$ with $\nu=1$ as determined by our collective coordinate approach (\ref{e.occ1})--(\ref{e.occ3}) (dotted lines, online red), the Lagrangian collective coordinate approach (\ref{e.lcc1})--(\ref{e.lcc2}) (dashed lines, online yellow). The continuous line (online blue) depicts results of a simulation of the full perturbed KdV equation (\ref{e.pkdv}).}
\label{f.l_nu1}
\end{figure}

In Figures~\ref{f.l_nu0p01} and \ref{f.l_nu1} we compare the predictions of the two different collective coordinate approaches to results from a numerical simulation of the partial differential equation (\ref{e.pkdv}) with the deterministic perturbation $P(u,x,t)=-\nu u$. We extract the collective coordinates from the simulation by a nonlinear least square fitting to solutions of the form (\ref{e.solitonA}). Figures~\ref{f.l_nu0p01} and \ref{f.l_nu1} show results for a small perturbation with $\nu=0.001$ and a larger perturbation with $\nu=1$, respectively. Interestingly, the results suggest that respecting the additional variational structure is advantageous for small perturbations, where the Lagrangian collective coordinate approach outperforms our Galerkin approximation based collective coordinate framework. Once the perturbations, however, are sufficiently large such that one cannot view the equation as a perturbed variational equation, the performance reverses and our approach becomes superior as clearly seen in Figure~\ref{f.l_nu1}. It is pertinent to mention that the Lagrangian approach is not able to capture the reversal in propagation experienced by the perturbed solitary wave (cf. (\ref{e.lcc2}), which our approach captures (cf. (\ref{e.occ3})), albeit too strongly. The discrepancy is caused, we suspect, by the solitary wave now being able to interact strongly with linear waves which is not captured by the collective coordinate approach.
	
%%%%%%%%%%%%%%%%%%%%%%%%%%%%%%%%%%%%%%%%%%%%%%%%%%%%%%%%%%%%%%%%%%
\section{Numerical scheme}\label{a.num}
Here we outline the numerical scheme used to solve \eqref{e.skdv}, based on the scheme given in \cite{Lord}. We use a spatial discretisation of size $\Delta x$ and a temporal discretisation of time $\Delta t$ with periodic boundary conditions. We denote the numerical solution at time $t^n=n\Delta t$ as  $U^n$, where $U^n=[U^n_0,U^n_1,\ldots,U^n_k,\ldots,U^n_{N} ]^T$ and $U^n_{k}=U(x_k,t^n)$ where $x_k=-L+k\Delta x$ and $N=2L/\Delta x +1$ denotes the number of spatial gridpoints. We use centred finite difference schemes that are second order in space for the operators $\partial_x$ and $-\partial_{xxx}$. The associated $N\times N$ matrices we denote by $\mathcal{D}$ and $\mathcal{L}$, respectively. We further define the vector $\mathcal{N}(U^n)=6U^n * \mathcal{D}U^n$, where the product is done element-wise. The multiplicative noise factor is either $R(U^n)=U^n$ or $R(U^n)=DU^n$.\\

To initialize, we employ a simple Euler-Maruyama step as the first time step with 
\begin{align*}
U^1=U^0+\Delta t\left(\mathcal{L}U^0+\mathcal{N}(U^0) \right) + \sigma R(U^0)\Delta W,
\end{align*}
where $\Delta W=\xi \sqrt{\Delta t}$ and $\xi\sim N(0,1)$. For all subsequent time steps, we solve the deterministic part by using a Crank-Nicolson method for the linear term and Adams-Bashforth for the nonlinear term, resulting in the scheme for $n>1$
\begin{align*}
U^{n+1}=\left(I-\frac{\Delta t}{2}\mathcal{L} \right)^{-1}\left[\left(I+\frac{\Delta t}{2}\mathcal{L} \right)U^n+\frac{\Delta t}{2}\left(3\mathcal{N}\left(U^n\right)-\mathcal{N}\left(U^{(n-1)}\right)\right)+\sigma R(U^n)\Delta W \right]. 
\end{align*}

%%%%%%%%%%%%%%%%%%%%%%%%%%%%%%%%%%%%%%%%%%%%%%%%%%%%%%%%%%%%%%%%%

\section{Perturbative collective coordinate ansatz}
\label{a.ccL}
Here we present results for the perturbative ansatz (\ref{e.uhatx}) for the stochastically perturbed KdV equation (\ref{e.skdv}) with $R(u)=u$, which we recall here 
\begin{align*}
\tilde u(x,t;{\bf{c}})=\hat u(x,t) + \alpha {\hat u}_x(x,t),
%\label{e.uhat_alphaux}
\end{align*}
with ${\bf{c}}=(\tilde\kappa,\tilde w,\tilde \phi)$. Note that the location of $\hat u$ is labelled here by $\tilde \phi$ as opposed to $\phi$ in the collective coordinate ansatz (\ref{e.uhat}); the additional odd function $\hat u_x$ leads to a shift in the position of the maximum of $\tilde u(x,t)$ with $\tilde \phi \neq \phi$. Similarly, the amplitudes and inverse widths are altered as well, and we label them here $\tilde \kappa$ and $\tilde w$. In addition to the collective coordinates $\kappa$, $w$ and $\phi$ this ansatz contains the collective coordinate $\alpha$. This implies an additional projection of the error onto $\partial \tilde u/\partial \alpha$ according to (\ref{e.cc_drift_gen})--(\ref{e.cc_noise_gen}). This yields the following rather unwieldy evolution equations for the collective coordinates
\begin{align}
\ud\tilde{\kappa} &=\frac{64 \tilde{\alpha}  \tilde{\kappa}  \tilde{w}^2 }{35 \left((240 \pi ^2 -1260) \tilde{\alpha} ^2 \tilde{w}^2+84 \pi ^2-805\right)}
\left(35 (\tilde{w}^2-\tilde{\kappa}) (15 + 4 \pi^2)  \right.
\nonumber
\\
&\qquad\left. + 4 \tilde{w}^2 ((\tilde{w}^2-\tilde{\kappa}) (100 \pi^2 -1309  ) + 784 \tilde{w}^2) \tilde{\alpha}^2\right)\dt +\sigma\tilde{\kappa}\dW,
\label{e.acc1}
\\
\ud \tilde{w}&=\frac{768\tilde{\alpha} \tilde{w}^3 \left(42 \tilde{\alpha} ^2 \tilde{\kappa}  \tilde{w}^2+25 (\tilde{w}^2-\tilde{\kappa}) \right)}{5 \left((240 \pi ^2 -1260)\tilde{\alpha} ^2 \tilde{w}^2+84 \pi ^2-805\right)}\dt,
\label{e.acc2}
\\
\ud\tilde \phi&=\frac{4}{ \left(240 \pi ^2-1260\right) \tilde{\alpha} ^2 \tilde{w}^2+84 \pi ^2-805}\left(4 \tilde{\alpha} ^2 \tilde{w}^2 \left(-4 \left(40 \pi ^2-259\right) (\tilde{w}^2-\tilde{\kappa})+\left(60 \pi ^2-511\right) \tilde{w}^2\right)\right.
\nonumber\\
&\qquad\left.-112 \left(2 \pi ^2-15\right) (\tilde{w}^2-\tilde{\kappa})+7 \left(12 \pi ^2-115\right) \tilde{w}^2\right)\dt,
\label{e.acc3}
\\
\ud\tilde{\alpha}&=\frac{16 }{7 \left((240 \pi ^2-1260) \tilde{\alpha} ^2 \tilde{w}^2+84 \pi ^2-805\right)}
\left(35 (\tilde{w}^2-\tilde{\kappa}) (15 - 4 \pi^2) -4 \tilde{w}^2\tilde{\alpha}^2((\tilde{w}^2-\tilde{\kappa}) ( 128 \pi^2-553)   \right.
\nonumber
\\
&\qquad\left. + 343 \tilde{w}^2)  -  80 \tilde{w}^4 \tilde{\alpha}^4((\tilde{w}^2-\tilde{\kappa}) ( 4 \pi^2-70) + 49 \tilde{w}^2)
 \right)\dt.
\label{e.acc4}
\end{align}

For completeness we list the evaluations of the projections involved in deriving (\ref{e.acc1})--(\ref{e.acc4}) 
\begin{align*}
&\langle u^2\rangle=\frac{16\tilde{\kappa}^2 \left(4 \tilde{\alpha} ^2 \tilde{w}^2+5\right)}{15 \tilde{w}}, 
\quad
\langle u\pdif{u}{\tilde{w}}\rangle=\frac{8 \tilde{\kappa}^2  \left(4 \tilde{\alpha} ^2 \tilde{w}^2-5\right)}{15 \tilde{w}^2},
\quad
\langle u\pdif{u}{\tilde{\alpha}}\rangle=\frac{64 \tilde{\alpha}  \tilde{\kappa}^2  \tilde{w}}{15},\\
&\langle\left(\pdif{u}{\tilde{w}}\right)^2\rangle=\frac{4 \tilde{\kappa} ^2 \left(16 \left(5 \pi ^2-21\right) \tilde{\alpha} ^2 \tilde{w}^2+28 \pi ^2\right)}{315 \tilde{w}^3},
\quad
\langle\pdif{u}{\tilde{w}}\pdif{u}{\tilde{\alpha}}\rangle=\frac{32 \tilde{\alpha}  \tilde{\kappa} ^2}{15},
\\
&\langle\left(\pdif{u}{x}\right)^2\rangle=\frac{64}{105} \tilde{\kappa} ^2 \tilde{w} \left(20 \tilde{\alpha} ^2 \tilde{w}^2+7\right),
\quad
\langle\pdif{u}{x}\pdif{u}{\tilde{\alpha}}\rangle=\frac{64 \tilde{\kappa} ^2 \tilde{w}}{15},
\quad
\langle\left(\pdif{u}{\tilde{\alpha}}\right)^2\rangle=\frac{64 \tilde{\kappa} ^2 \tilde{w}}{15},
\\
&\langle u\pdif{u}{x}\pdif{u}{\tilde{w}}\rangle=\frac{1024}{225} \tilde{\alpha} ^3 \tilde{\kappa} ^3 \tilde{w}^2,
\quad
\langle u\left(\pdif{u}{x}\right)^2\rangle=-\frac{512}{105} \tilde{\kappa} ^3 \tilde{w} \left(4 \tilde{\alpha} ^2 \tilde{w}^2+1\right),
\quad
\langle u\pdif{u}{x}\pdif{u}{\tilde{\alpha}}\rangle=-\frac{512}{105}  \tilde{\kappa} ^3 \tilde{w},
\\
&\langle \pdif{^3u}{x^3}\pdif{u}{x} \rangle=-\frac{256}{105} \tilde{\kappa} ^2 \tilde{w}^3 \left(28 \tilde{\alpha} ^2 \tilde{w}^2+5\right),
\quad
\langle \pdif{^3u}{x^3}\pdif{u}{\tilde{\alpha}}\rangle =-\frac{256}{21} \tilde{\kappa} ^2 \tilde{w}^3,
\\
&\langle u\pdif{u}{x}\rangle=\langle\pdif{u}{\tilde{w}}\pdif{u}{x}\rangle=\langle u^2\pdif{u}{x}\rangle=\langle u\pdif{^3u}{x^3}\rangle=\langle \pdif{^3u}{x^3}\pdif{u}{\tilde{w}}\rangle=0.
\end{align*}

For $|\alpha|\ll 1$, the evolution equations for the collective coordinates (\ref{e.acc1})--(\ref{e.acc4}) become upon neglecting terms of $\mathcal{O}(\alpha^2)$, 
\begin{align}
\ud\tilde{\kappa}&=\sigma\tilde{\kappa}\dW + \tilde{\alpha} \frac{64 \left(15+4 \pi ^2\right) }{84 \pi ^2-805}\tilde{\kappa}  \tilde{w}^2 \left(\tilde{w}^2-\tilde{\kappa} \right)\dt,\label{e.ans2_k}\\
\ud \tilde{w}&=\frac{3840}{84 \pi ^2-805} \tilde{\alpha}  \tilde{w}^3\left(\tilde{w}^2-\tilde{\kappa} \right)\dt,\label{e.ans2_w}\\
\ud \tilde{\phi}&=\left(4\tilde{w}^2-\frac{64 \left(2 \pi ^2-15\right)}{12 \pi ^2-115} (\tilde{w}^2-\tilde{\kappa})\right)\dt,\label{e.ans2_p}\\
\ud\tilde{\alpha}&=\frac{80 \left(4 \pi ^2-15\right)}{805-84 \pi ^2}\left(\tilde{w}^2-\tilde{\kappa} \right)\dt. \label{e.ans2_a}
\end{align}
It is readily seen that the magnitude $\tilde{\alpha}(t)$ of the perturbation to the $\sech^2$-profile of the solitary wave $\hat u(x,t)$ is proportional to $\delta(t)/w$ (cf. (\ref{e.delta})). Comparing with (\ref{e.u_dphi}), we obtain as expected for a Taylor expansion in the location, that the location $\phi$ of the $\sech^2$-profile $\hat u(x,t)$ is recovered by $\phi = \tilde \phi - \tilde{\alpha}$. \\

\begin{figure}[htbp]
	\centering
	\includegraphics[width = 0.4\columnwidth]{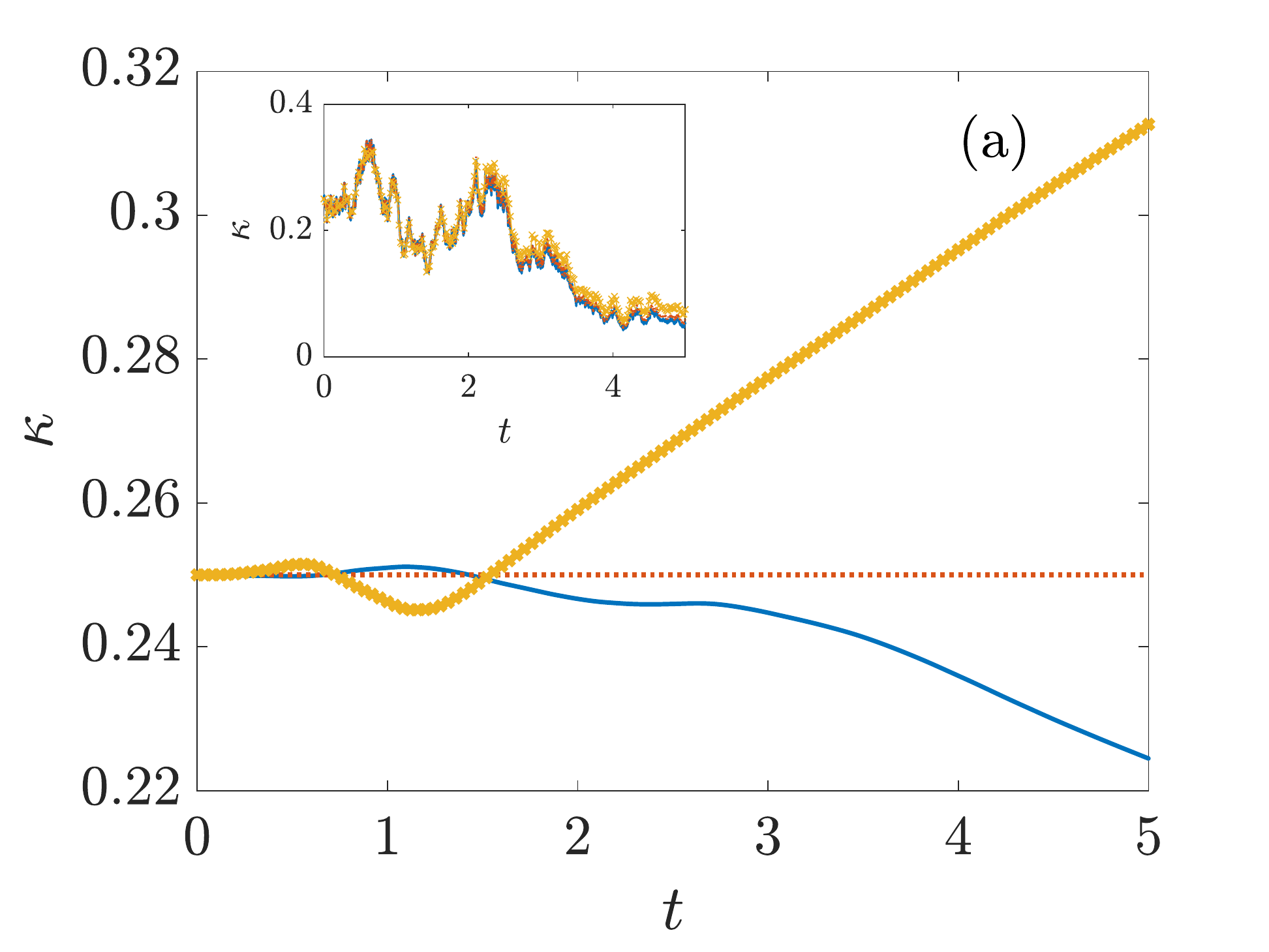}
	\includegraphics[width = 0.4\columnwidth]{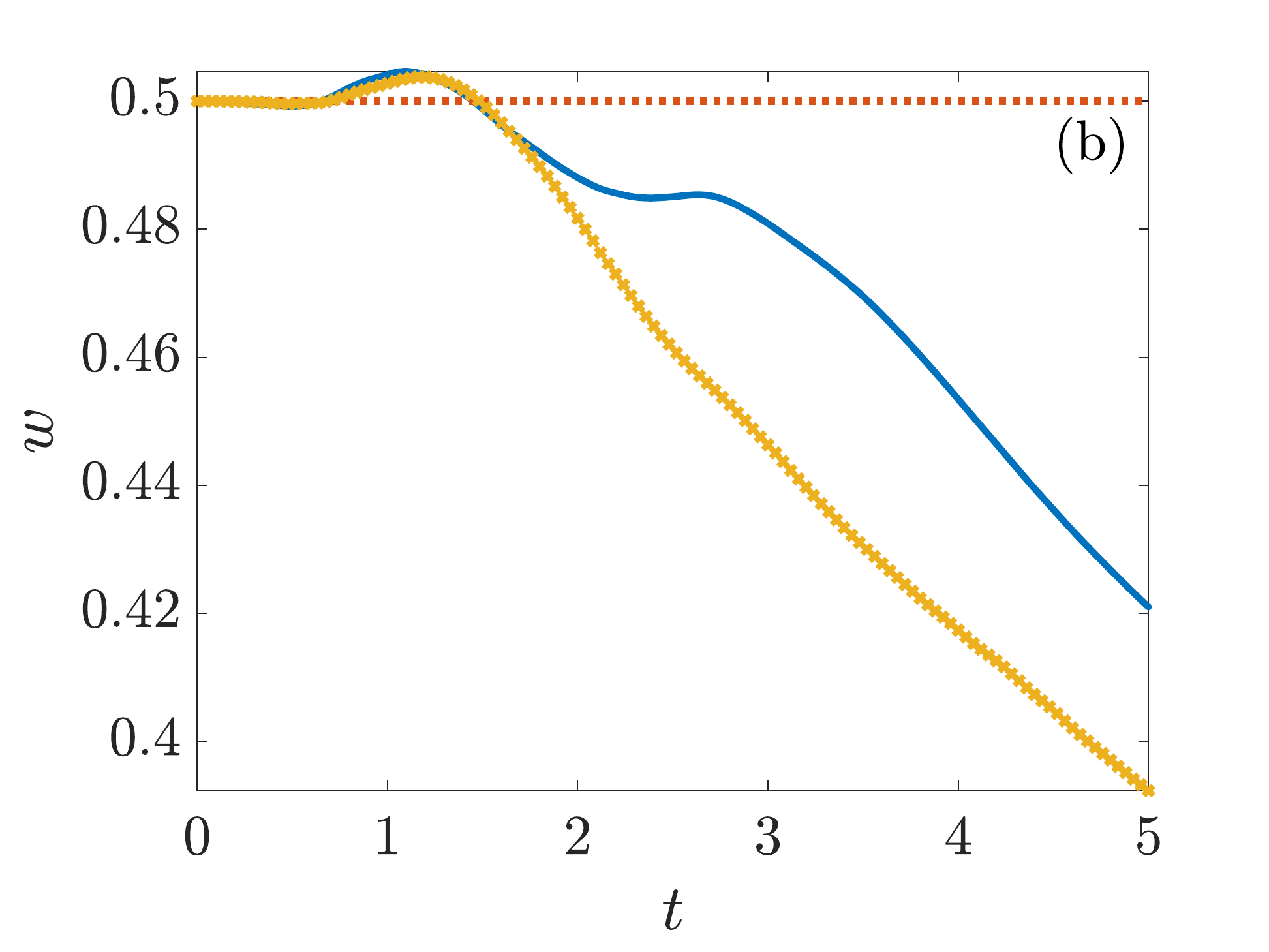}\\		
	\includegraphics[width = 0.4\columnwidth]{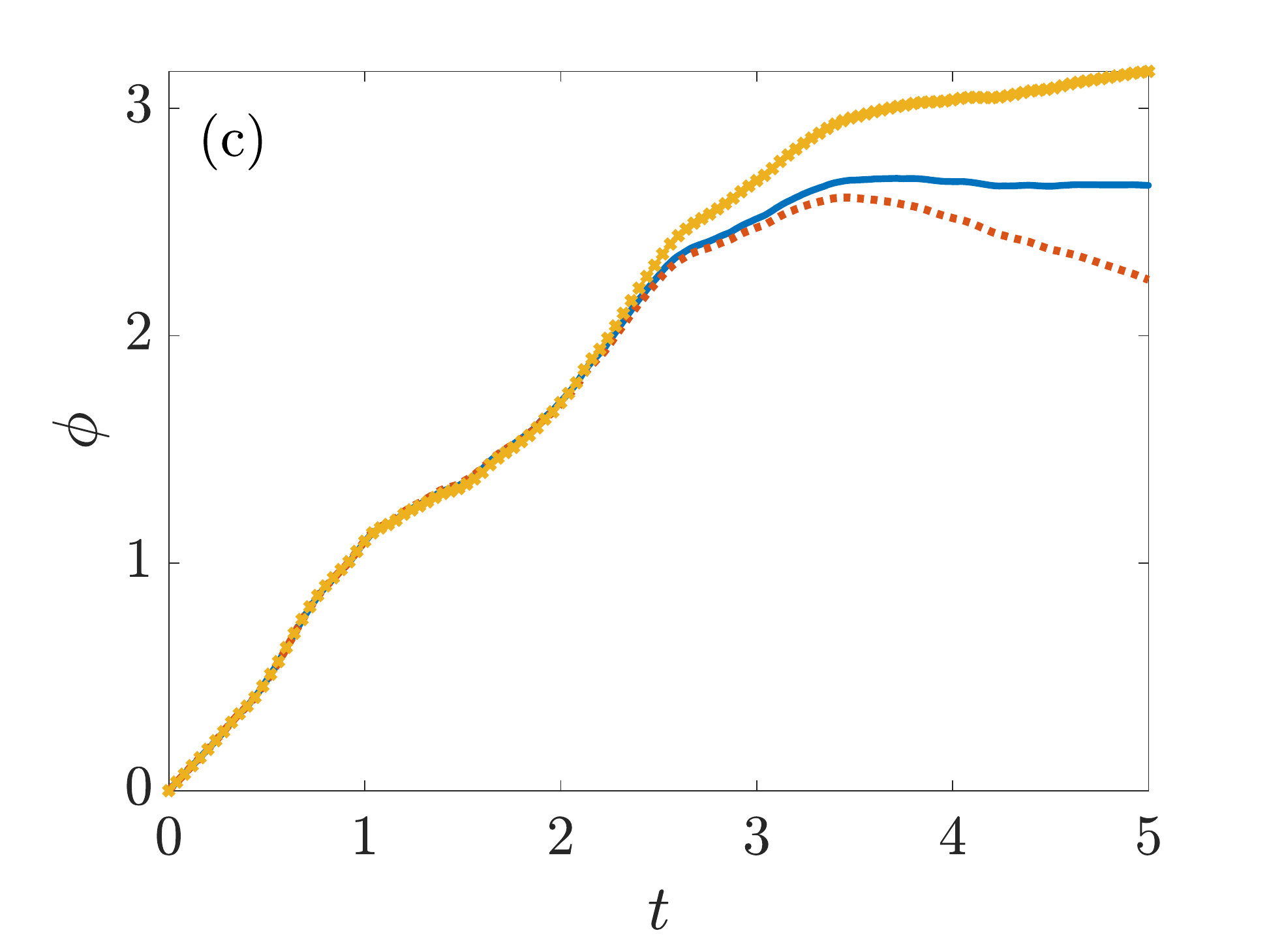}
	\includegraphics[width = 0.4\columnwidth]{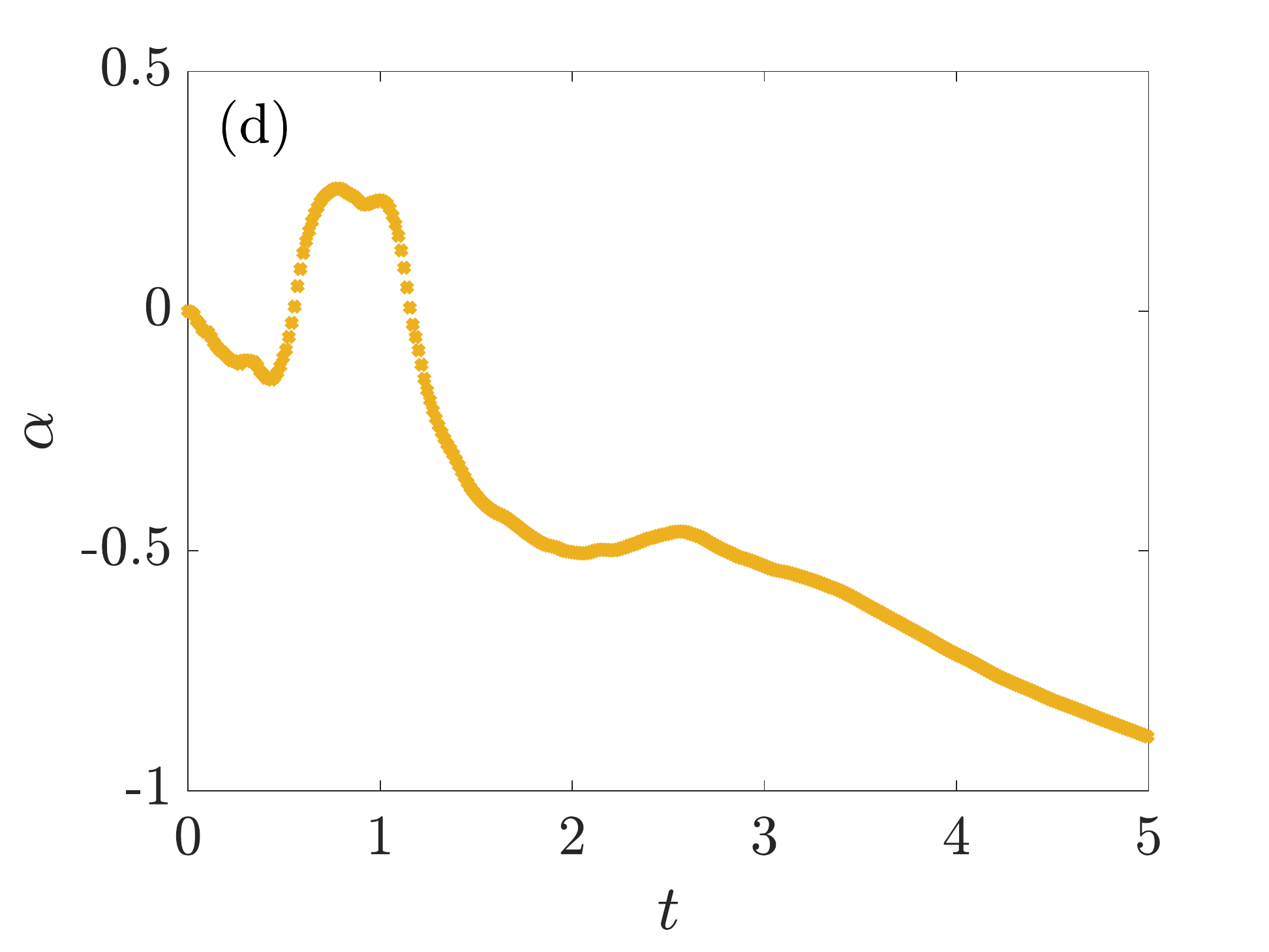}
\caption{Amplitude $\kappa$, inverse width $w$, location $\phi$ and $\alpha$ of a solitary wave ansatz function (\ref{e.uhatx}) for the stochastically perturbed KdV equation (\ref{e.u}) as a function of time. Continuous lines (online blue) are obtained from a direct simulation of the stochastically perturbed Korteweg-de Vries equation (\ref{e.u}). Dotted lines (online red) are obtained from the collective coordinate approach (\ref{e.u_cc1})--(\ref{e.u_cc3}) (ignoring $\alpha$). Crosses (online yellow) are obtained from the perturbative collective coordinate approach (\ref{e.acc1})-(\ref{e.acc4}), where the corresponding values for $\kappa$, $w$ and $\phi$ are obtained by a nonlinear least square fit of the function (\ref{e.uhatx}) to the $\sech^2$-function (\ref{e.uhat}). %For clarity, only every eightieth data point is plotted for $\kappa$ obtained from (\ref{e.acc1}). 
The parameters used for the simulation are the same as in Fig~\ref{f.u_cc_t}.}
\label{f.ccL}
\end{figure}

In Figure~\ref{f.ccL} we show numerical results of the perturbative collective coordinate equations  (\ref{e.acc1})--(\ref{e.acc4})  and compare them with the results of the collective coordinate without including the perturbative term with amplitude $\tilde{\alpha}$, e.g. (\ref{e.u_cc1})--(\ref{e.u_cc3}). Note that in the perturbative ansatz (\ref{e.uhatx}) the location of the solitary wave is not given by $x=\tilde \phi$ as in the original collective coordinate ansatz  (\ref{e.u_cc1})--(\ref{e.u_cc3}); the additional term $\alpha\hat u_x$ leads to a shift of the maximum. Therefore to compare with the full solution we perform at each time a nonlinear least square fit of the ansatz function (\ref{e.uhatx}) to the $\sech^2$-function (\ref{e.uhat}). The perturbative collective coordinates closely follow the evolution of the coherent solitary wave up to times $t\approx 1.8$.  The amplitude $\kappa$ is well reproduced by both collective coordinates even for times $t>1.8$ (see inset). However, this is due to the overall multiplying factor $\mu(t)$ of the geometric Brownian motion. If the factor is removed, then $\kappa(t)/\mu(t)$ shows more clearly the differences between the two collective coordinate approaches and the corresponding solution of the stochastic SPDE (\ref{e.u}). In particular, the collective coordinate approach (\ref{e.acc1})-(\ref{e.acc4}) deviates strongly for $t>2$. Whereas in the original collective coordinate ansatz the inverse width is constant in time (cf.  (\ref{e.u_cc2})), $\tilde{w}$ is now temporally varying and better captures the dynamics of the full SPDE. Note that $w$ deviates from the corresponding shape variable $w$ of the SPDE solution around $t=t^\star$. The corresponding values for the coherence times are $\tau_c=2.83$ and $t^\star=2.23$.

%%%%%%%%%%%%%%%%%%%%%%%%%%%%%%%%%%%%%%%%%%%%%%%%%%%%%%%%%%%%%%%%%
	
%\bibliographystyle{siamplain}
\bibliographystyle{abbrvnat}
%\bibliographystyle{apsrev4-1}
%\bibliography{skdv}

\end{document}